\documentclass[12pt,preprint]{aastex}

\shorttitle{NGC 5986}
\shortauthors{Johnson et al.}

\newcommand\iso[2]{$^{\rm #1}$#2}

\begin{document}

\title{Chemical Complexity in the Eu--enhanced Monometallic Globular Cluster
NGC 5986\footnote{This paper includes data gathered with the 6.5m 
\emph{Magellan} Telescopes located as Las Campanas Observatory, Chile.}}

\author{
Christian I. Johnson\altaffilmark{1,2},
Nelson Caldwell\altaffilmark{1},
R. Michael Rich\altaffilmark{3},
Mario Mateo\altaffilmark{4},
John I. Bailey, III\altaffilmark{5},
Edward W. Olszewski\altaffilmark{6}, and
Matthew G. Walker\altaffilmark{7}
}

\altaffiltext{1}{Harvard--Smithsonian Center for Astrophysics, 60 Garden
Street, MS--15, Cambridge, MA 02138, USA; cjohnson@cfa.harvard.edu; 
ncaldwell@cfa.harvard.edu}

\altaffiltext{2}{Clay Fellow}

\altaffiltext{3}{Department of Physics and Astronomy, UCLA, 430 Portola Plaza,
Box 951547, Los Angeles, CA 90095-1547, USA; rmr@astro.ucla.edu}

\altaffiltext{4}{Department of Astronomy, University of Michigan, Ann Arbor,
MI 48109, USA; mmateo@umich.edu}

\altaffiltext{5}{Leiden Observatory, Leiden University, P. O. Box 9513, 2300RA 
Leiden, The Netherlands; baileyji@strw.leidenuniv.nl}

\altaffiltext{6}{Steward Observatory, The University of Arizona, 933 N. Cherry
Avenue, Tucson, AZ 85721, USA; eolszewski@as.arizona.edu}

\altaffiltext{7}{McWilliams Center for Cosmology, Department of Physics, 
Carnegie Mellon University, 5000 Forbes Avenue, Pittsburgh, PA 15213, USA; 
mgwalker@andrew.cmu.edu}

\begin{abstract}

NGC 5986 is a poorly studied but relatively massive Galactic globular cluster
that shares several physical and morphological characteristics with 
``iron--complex" clusters known to exhibit significant metallicity and heavy 
element dispersions.  In order to determine if NGC 5986 joins the iron--complex
cluster class, we investigated the chemical composition of 25 red giant branch 
and asymptotic giant branch cluster stars using high resolution spectra 
obtained with the \emph{Magellan}--M2FS instrument.  Cluster membership was 
verified using a combination of radial velocity and [Fe/H] measurements, and 
we found the cluster to have a mean heliocentric radial velocity of $+$99.76 km
s$^{\rm -1}$ ($\sigma$ = 7.44 km s$^{\rm -1}$).  We derived a mean metallicity
of [Fe/H] = --1.54 dex ($\sigma$ = 0.08 dex), but the cluster's small 
dispersion in [Fe/H] and low [La/Eu] abundance preclude it from being an 
iron--complex cluster.  NGC 5986 has $\langle$[Eu/Fe]$\rangle$ = $+$0.76 dex 
($\sigma$ = 0.08 dex), which is among the highest ratios detected in a Galactic
cluster, but the small [Eu/Fe] dispersion is puzzling because such high values 
near [Fe/H] $\sim$ --1.5 are typically only found in dwarf galaxies exhibiting 
large [Eu/Fe] variations.  NGC 5986 exhibits classical globular cluster 
characteristics, such as uniformly enhanced [$\alpha$/Fe] ratios, a small 
dispersion in Fe--peak abundances, and (anti--)correlated light element 
variations.  Similar to NGC 2808, we find evidence that NGC 5986 may host at 
least 4--5 populations with distinct light element compositions, and the 
presence of a clear Mg--Al anti--correlation along with an Al--Si correlation 
suggests that the cluster gas experienced processing at temperatures $\ga$ 
65--70 MK.  However, the current data do not support burning temperatures 
exceeding $\sim$100 MK.  We find some evidence that the first and second 
generation stars in NGC 5986 may be fully spatially mixed, which could indicate
that the cluster has lost a significant fraction of its original mass.

\end{abstract}


\section{INTRODUCTION}

In contrast to the expectation that stars within a single globular cluster 
should exhibit relatively uniform composition patterns, early high resolution
spectroscopic analyses found that the abundances of elements such as O, Na, and
Al can vary significantly between stars in the same cluster (e.g., Cohen 1978; 
Peterson 1980; Norris et al. 1981; Norris \& Pilachowski 1985; Hatzes 1987).
Subsequent observations confirmed these early results and expanded upon the
realization that not only are certain element pairs, such as O and Na, 
(anti--)correlated (e.g., Drake et al. 1992; Kraft et al. 1993, 1997;
Norris \& Da Costa 1995; Pilachowski et al. 1996; Sneden et al. 1997, 2000;
Ivans et al. 1999, 2001; Cavallo \& Nagar 2000), but that similar light element
abundance patterns are present in nearly all old ($\ga$ 6--8 Gyr) Galactic 
globular clusters (e.g., Carretta et al. 2009a,b; see also reviews by Gratton 
et al. 2004, 2012a).  Recently, the list has grown to include old extragalactic
globular clusters as well (Mucciarelli et al. 2009; Schiavon et al. 2013; 
Larsen et al. 2014; Dalessandro et al. 2016; Niederhofer et al. 2017; Hollyhead
et al. 2017), which supports the suggestion by Carretta et al. (2010a) that 
a population's stars must exhibit an O--Na anti--correlation to be labeled as 
a globular cluster.

Since the heavier $\alpha$ and Fe--peak element abundances typically
exhibit small ($\la$ 0.1 dex) star--to--star dispersions within globular 
clusters, the light element (anti--)correlations have been interpreted as a 
by--product of high temperature ($>$ 40 MK) proton--capture burning (e.g., 
Denisenkov \& Denisenkova 1990; Langer et al. 1993, 1997; Prantzos et al. 
2007).  Initial analyses of bright red giant branch (RGB) stars suggested that 
\emph{in situ} processing and deep mixing could be responsible for a majority 
of the abundance variations (e.g., see review by Kraft 1994).  However, the 
discovery of similar chemical trends in globular cluster main--sequence and 
subgiant branch stars (e.g., Briley et al. 1994, 1996; Gratton et al. 2001; 
Cohen \& Mel{\'e}ndez 2005; Bragaglia et al. 2010a; D'Orazi et al. 2010; 
Dobrovolskas et al. 2014) revealed that the chemical composition variations 
must be a result of pollution from a previous generation of more massive stars.

Interestingly, the introduction of \emph{Hubble Space Telescope} (\emph{HST})
optical and near--UV photometry to the field showed that globular clusters form
distinct populations of stars with unique light element compositions (e.g., 
Piotto et al. 2007, 2012, 2015; Milone et al. 2012a, 2012b) rather than the 
continuous distributions that are expected from simple dilution models.  When 
combined with ground--based high resolution spectroscopy, combinations of color
and pseudo--color indices can be used to create ``chromosome" maps (Milone et 
al. 2015a, 2015b) that anchor the photometry of different stellar populations to
specific compositions.  Furthermore, a combination of these chromosome maps, 
isochrone fitting, and direct measurements have shown that, as a consequence of 
proton--capture nucleosynthesis, stars with enhanced abundances of 
[N/Fe]\footnote{[A/B] $\equiv$ log(N$_{\rm A}$/N$_{\rm B}$)$_{\rm star}$ --
log(N$_{\rm A}$/N$_{\rm B}$)$_{\sun}$ and log $\epsilon$(A) $\equiv$
log(N$_{\rm A}$/N$_{\rm H}$) + 12.0 for elements A and B.}, 
[Na/Fe], and [Al/Fe] and low abundances of [C/Fe], [O/Fe], and [Mg/Fe] are 
also enriched in He with $\Delta$Y ranging from $\sim$ 0.01--0.15 (e.g., Norris
2004; Piotto et al. 2005; Bragaglia et al. 2010a, 2010b; Dupree et al. 2011;
Pasquini et al. 2011; Milone et al. 2012a; Villanova et al. 2012; Milone 2015).
In fact, the discrete nature of multiple populations in globular clusters, 
the variety of He enhancements and light element patterns, and the large 
fraction ($\sim$60--80$\%$; Carretta 2015, their Figure 16) of polluted 
``second generation" stars in clusters place strong constraints on the possible
pollution sources and enrichment time scales.  However, identifying and 
quantifying the exact pollution source(s) remain unsolved problems (e.g., see 
critical discussions in Renzini 2008; Valcarce \& Catelan 2011; Bastian et al. 
2015; Bastian \& Lardo 2015; Renzini et al. 2015; D'Antona et al. 2016).

As an added complication, a small but growing number of $\sim$10
``iron--complex" globular clusters are now known to possess intrinsic spreads 
in [Fe/H] that are found concurrent with the aforementioned light element 
abundance variations (e.g., Marino et al. 2009, 2011a, 2011b, 2015; Carretta et
al. 2010b, 2011; Johnson \& Pilachowski 2010; Yong et al. 2014a; Johnson et al. 
2015a, 2017)\footnote{Note that the metallicity spreads for some clusters are 
disputed (Mucciarelli et al. 2015a, 2015b; Lardo et al. 2016; but see also Lee 
2016).}.  These clusters are suspected to be the remnant cores of former dwarf 
spheroidal galaxies (e.g., Bekki \& Freeman 2003; Lee et al. 2007; Georgiev et 
al. 2009; Da Costa 2016), and therefore may trace a part of the Galaxy's minor 
merger history.  Despite exhibiting broad ranges in mass, metallicity, and 
galactocentric distance, iron--complex clusters share several notable 
features: (1) they are among the most massive clusters in the Galaxy and all 
have M$_{\rm V}$ $<$ --8.3; (2) most have very blue and extended 
horizontal branch morphologies; (3) the dispersion in [Fe/H] is $\ga$ 0.1 dex
when measured from high resolution spectra; (4) several clusters contain 
discrete metallicity groups rather than just broadened distributions; (5) many
have $\langle$[Fe/H]$\rangle$ $\sim$ --1.7; (6) and in all cases the stars with
higher [Fe/H] have strong s--process enhancements.  Using these characteristics
as a template, we can search for new iron--complex clusters by measuring light
and heavy element abundances in previously unobserved massive clusters with
extended blue horizontal branches.

In this context, Da Costa (2016) noted that NGC 5986 may be a promising 
iron--complex candidate.  This cluster is relatively massive with 
M$_{\rm V}$ = --8.44 (Harris 1996; 2010 revision), has an irregular
and highly eccentric prograde--retrograde orbit (Casetti--Dinescu et al. 2007; 
Allen et al. 2008; Moreno et al. 2014), hosts a predominantly blue and 
very extended horizontal branch (Kravtsov et al. 1997; Ortolani et al. 2000;
Rosenberg et al. 2000; Alves et al. 2001; Momany et al. 2004; Piotto et al. 
2015), and is estimated to have [Fe/H] $\approx$ --1.6 (Zinn \& West 1984; 
Geisler et al. 1997; Ortolani et al. 2000; Kraft \& Ivans 2003; Jasniewicz et 
al. 2004; Rakos \& Schombert 2005; Dotter et al. 2010).  However, except for
an investigation into the composition of two post--asymptotic giant branch 
(AGB) stars by Jasniewicz et al. (2004), no detailed chemical abundance 
analysis has been performed for NGC 5986.  Therefore, we present here a 
detailed composition analysis of 25 RGB and AGB stars in NGC 5986, and
aim to determine if the cluster belongs to the iron--complex class or is 
instead a more typical monometallic cluster.

\section{OBSERVATIONS AND DATA REDUCTION}

The spectra for this project were acquired using the Michigan--\emph{Magellan} 
Fiber System (M2FS; Mateo et al. 2012) and MSpec spectrograph mounted on the 
Clay--\emph{Magellan} 6.5m telescope at Las Campanas Observatory.  The 
observations were obtained on 2016 June 22 and 2016 June 27 in clear weather 
and with median seeing ranging between about 0.9$\arcsec$ and 1.2$\arcsec$.  
All observations utilized the same instrument configuration that included 
binning both CCDs 1 $\times$ 2 (dispersion $\times$ spatial) with four 
amplifiers in a slow readout mode.  The ``red" and ``blue" spectrographs were 
each configured in high resolution mode, and the 1.2$\arcsec$ fibers, 125$\mu$m
slits, and echelle gratings produced a typical resolving power of R $\equiv$ 
$\lambda$/$\Delta$$\lambda$ $\approx$ 27,000.  We also employed the 
``Bulge$\_$GC1" order blocking filters that provide 6 consecutive orders
spanning $\sim$6120--6720 \AA\, at the cost of only using 48 of the possible 
256 fibers.

\begin{figure}
\epsscale{1.00}
\plotone{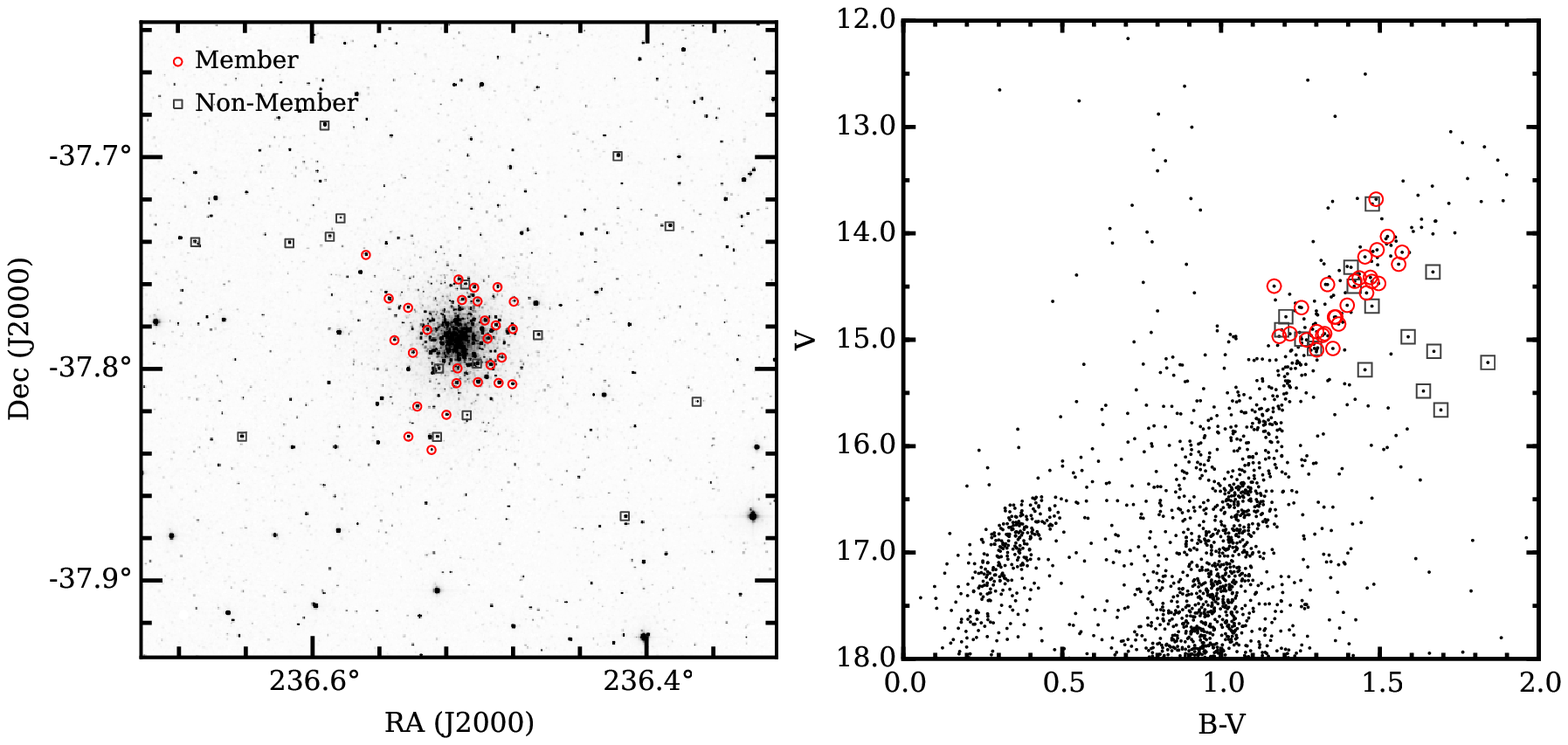}
\caption{\emph{Left:} a 2MASS (Skrutskie et al. 2006) J--band image of NGC 5986
is shown.  The radial velocity members and non--members identified in this work
are indicated by open red circles and open grey boxes, respectively.
\emph{Right:} a V versus B--V color--magnitude diagram from Alves et al. (2001)
is shown with the same member and non--member stars identified.}
\label{f1}
\end{figure}

Potential target stars were identified using J and K$_{\rm S}$ photometry 
available from the Two Micron All Sky Survey (2MASS; Skrutskie et al. 2006).  
We selected stars with J--K$_{\rm S}$ colors of 0.7--1.1 magnitudes and
K$_{\rm S}$ ranging from 10.0--11.8 magnitudes.  These selection criteria are
equivalent to a range of about 1.1--1.9 magnitudes in B--V color and 13.5--15.7 
magnitudes in the V--band, which we illustrate in Figure \ref{f1} using 
optical photometry from Alves et al. (2001).  A broad color range was 
adopted in order to probe the possible existence of intrinsic metallicity
variations in the cluster.  However, Figure \ref{f1} shows that all of the 
stars redder than the formal RGB were determined to have radial velocities
inconsistent with cluster membership (see Section 3).  We note also that two
stars with high membership probabilities (2MASS 15460024--3748232 and 2MASS
15460078--3745426) may be bluer and brighter than the formal RGB and AGB
sequences, and as a consequence could be post--AGB stars.

The coordinates for all targets were obtained from the 2MASS catalog, and we 
were able to place fibers on 43 stars and 5 sky positions using a single
configuration.  Fibers were assigned to targets ranging from about
0.8--8$\arcmin$ from the cluster center, but the member stars were only found
inside $\sim$3.5$\arcmin$.  Stars inside $\sim$0.8$\arcmin$ were avoided in 
order to mitigate the effects of blending and scattered light.  With respect to
the Alves et al. (2001) observations, we observed approximately 18$\%$ of all 
possible cluster stars with B--V in the range 1.1--1.9 magnitudes and with V 
between 13.5 and 16.0 magnitudes.  The star names, coordinates, and photometry 
for all target stars are provided in Table 1.  The evolutionary state, based on
a visual inspection of Figure \ref{f1}, is also provided in Table 1 for the 
member stars.

\subsection{Data Reduction}

The data reduction procedure followed the methods outlined by Johnson et al.
(2015b) in which the IRAF\footnote{IRAF is distributed by the National 
Optical Astronomy Observatory, which is operated by the Association of 
Universities for Research in Astronomy, Inc., under cooperative agreement with 
the National Science Foundation.} tasks \emph{CCDPROC}, \emph{zerocombine}, and
\emph{darkcombine} were used to trim the overscan regions, create master bias
and dark frames, and remove the bias and dark current effects.  These basic 
data reduction tasks were performed independently on each amplifier frame.  The
reduced images were then rotated and transposed using the \emph{imtranspose} 
task and combined via the \emph{imjoin} routine to create one full monolithic 
image per CCD per exposure.  

The remaining tasks of aperture tracing, flat--field correcting, scattered 
light removal, wavelength calibration, cosmic ray cleaning, and spectrum 
extraction were carried out using the \emph{dohydra} task.  Master sky spectra 
for each exposure set were created by scaling and combining the extracted sky 
fiber spectra, which were then subtracted from the object exposures.  The final
sky subtracted images were continuum normalized and combined after removing 
the heliocentric velocities from each exposure and dividing by a high 
signal--to--noise (S/N) telluric spectrum.  The final combined spectra had 
typical S/N ratios of approximately 100--200 per resolution element.

\section{RADIAL VELOCITIES AND CLUSTER MEMBERSHIP}

The radial velocities for all stars were calculated using the XCSAO (Kurtz \&
Mink 1998) cross--correlation routine.  The synthetic spectrum of a cool
metal--poor giant, smoothed and resampled to match the observations, was used 
as the reference template, and a heliocentric radial velocity value was 
independently determined for every order of each exposure.  However, we 
avoided regions with very strong lines (e.g., H$\alpha$) and those where 
residual telluric features may still be present (e.g., 6270--6320 \AA).  The
heliocentric radial velocity values listed in Table 1 represent the average 
velocity measurements of each order and exposure per star.  Similarly, the 
velocity error values in Table 1 represent the standard deviation of all 
radial velocity measurements for each star.  The average measurement 
uncertainties in Table 1 are 0.27 km s$^{\rm -1}$ ($\sigma$ = 0.07 km 
s$^{\rm -1}$) for the cluster members and 0.51 km s$^{\rm -1}$ ($\sigma$ = 0.27
km s$^{\rm -1}$) for the non--members.  

Despite being a relatively massive cluster, very little kinematic information
is available for NGC 5986.  Previous work estimated the systemic heliocentric
radial velocity of NGC 5986 to be $\sim$$+$90--97 km s$^{\rm -1}$ with a 
dispersion of $\sim$6--8 km s$^{\rm -1}$ (Hesser et al. 1986; Rutledge et al.
1997; Jasniewicz et al. 2004; Moni Bidin et al. 2009).  We find in general
agreement with past work but measure a higher heliocentric radial velocity of 
$+$99.76 km s$^{\rm -1}$ for NGC 5986 and a velocity dispersion of 7.44 km 
s$^{\rm -1}$.  For the non--member stars, we measure an average heliocentric 
radial velocity of --0.48 km s$^{\rm -1}$ ($\sigma$ = 60.24 km s$^{\rm -1}$).  

\begin{figure}
\epsscale{0.75}
\plotone{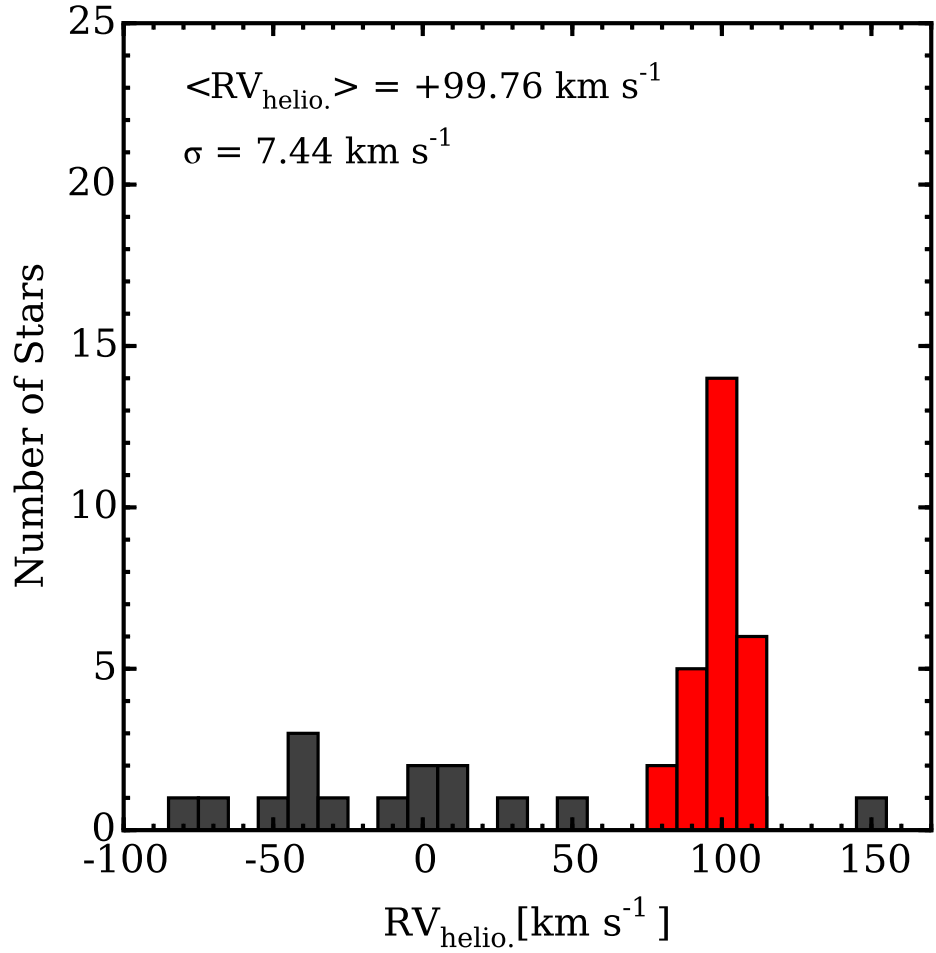}
\caption{The heliocentric radial velocity distribution of all stars observed
here is shown with bin sizes of 10 km s$^{\rm -1}$.  The radial velocity
members are identified by the red bins and the non--members are identified by
the grey bins.}
\label{f2}
\end{figure}

As can be seen in Figure \ref{f2}, the cluster and field star populations have 
clearly distinct velocity distributions.  However, we note that the star
2MASS 15455164--3747031 is likely a foreground interloper.  This star has a 
velocity of $+$105.81 km s$^{\rm -1}$, which is nominally consistent with 
cluster membership, but is significantly redder than the fiducial RGB sequence
shown in Figure \ref{f1}.  Therefore, we have classified 
2MASS 15455164--3747031 as a non--member in Table 1 and do not consider it 
further.  Using the final membership assignments outlined in Table 1, we find 
63$\%$ (27/43 stars) of the stars in our sample to be likely cluster members.

\section{ANALYSIS}

\subsection{Stellar Parameters and Abundance Determinations}

The analysis procedure adopted here closely follows the methods outlined in
Johnson et al. (2015a), and includes use of the same: line lists, reference 
Solar abundance ratios, equivalent width (EW) measuring software, grid of
$\alpha$--enhanced ATLAS9 model atmospheres (Castelli \& Kurucz 
2004)\footnote{The model atmosphere grid can be downloaded from: 
http://wwwuser.oats.inaf.it/castelli/grids.html.}, and local thermodynamic
equilibrium (LTE) line analysis code MOOG\footnote{The MOOG
source code is available at: http://www.as.utexas.edu/$\sim$chris/moog.html.} 
(Sneden 1973; 2014 version).  On average, we measured approximately 40 
\ion{Fe}{1} and 5 \ion{Fe}{2} lines per star, and used the EW values of both 
species and the \emph{abfind} driver in MOOG to iteratively solve for the model
atmosphere parameters effective temperature (T$_{\rm eff}$), surface gravity 
(log(g)), metallicity ([M/H]), and microturbulence ($\xi$$_{\rm mic.}$).  In 
particular, the T$_{\rm eff}$ values were determined by removing trends in 
plots of log $\epsilon$(\ion{Fe}{1}) versus excitation potential, log(g) was 
estimated by enforcing ionization equilibrium between \ion{Fe}{1} and 
\ion{Fe}{2}, $\xi$$_{\rm mic.}$ was set by removing trends in plots of 
log $\epsilon$(\ion{Fe}{1}) versus log(EW/$\lambda$), and the model metallicity
was set to the measured [Fe/H] abundance.  A list of the adopted model 
atmosphere parameters for each star is provide in Table 2.  Note that for 
2MASS 15455531--3748266 and 2MASS 15460957--3747333 we were unable to converge 
to a stable model atmosphere solution and do not consider these stars further. 

Similar to the case of \ion{Fe}{1} and \ion{Fe}{2}, the abundances of 
\ion{Si}{1}, \ion{Ca}{1}, \ion{Cr}{1}, and \ion{Ni}{1} were determined by an
EW analysis using the MOOG \emph{abfind} driver, the model atmosphere 
parameters listed in Table 2, and the line list provided in Johnson et al. 
(2015a; their Table 2).  On average, the \ion{Si}{1}, \ion{Ca}{1}, \ion{Cr}{1},
and \ion{Ni}{1} abundances were based on the measurements of 4, 6, 2, and 5 
lines, respectively.  All abundances have been measured relative to the 
metal--poor giant Arcturus, which is done to help offset effects due to 
departures from LTE and 1D versus 3D model atmosphere deficiencies.  The final
[Si/Fe], [Ca/Fe], [Cr/Fe], and [Ni/Fe] abundances for all member stars are
provided in Tables 3--4.

For \ion{O}{1}, \ion{Na}{1}, \ion{Mg}{1}, \ion{Al}{1}, \ion{La}{2}, and 
\ion{Eu}{2}, the abundances have been determined via the \emph{synth} spectrum
synthesis module in MOOG.  Similar to the EW analysis, the atomic and molecular
line lists within 10 \AA\ of each feature have been tuned to reproduce the 
Arcturus spectrum.  Specifically, the log gf values and reference Solar and 
Arcturus abundances for all species, except \ion{O}{1}, are the same as those 
in Johnson et al. (2015a).  Isotopic shifts and/or hyperfine broadening were
accounted for with \ion{La}{2} and \ion{Eu}{2} using the line lists from 
Lawler et al. (2001a,b).  For oxygen, we used the 6300.3 \AA\ [\ion{O}{1}] 
line and adopted the same atomic parameters and reference abundances as those 
provided by Johnson et al. (2014; their Table 2).  Although Dupree et al.
(2016) showed that the 6300.3 \AA\ feature can be affected by a star's 
chromosphere, the present data set does not provide enough information to 
reliably constrain a chromospheric model.  Therefore, the oxygen abundances
presented here are based only on radiative/convective equilibrium models.

All elements measured via spectrum synthesis included the updated CN line 
lists from Sneden et al. (2014), and the local CN lines were fit by fixing
[C/Fe] = --0.3, holding [O/Fe] at the best--fit value determined from the 
6300.3 \AA\ line, and treating the N abundance as a free parameter.  The 6319 
\AA\ \ion{Mg}{1} triplet required additional care because the lines are 
relatively weak and can be affected by a broad \ion{Ca}{1} auto--ionization 
feature.  We modeled the impact of the auto--ionization line by artificially 
modifying the log $\epsilon$(Ca) abundance during each synthesis such that the 
slope of the continuum around the \ion{Mg}{1} lines was well--fit.  The final 
[O/Fe], [Na/Fe], [Mg/Fe], [Al/Fe], [La/Fe], and [Eu/Fe] abundances for all 
member stars are provided in Tables 3--4, but are based on an average of only 
1--2 lines for each element.

\subsection{Model Atmosphere Parameter and Abundance Uncertainties}

We investigated uncertainties in the model atmosphere parameters using 
comparisons between spectroscopic and photometric T$_{\rm eff}$ and log(g) 
values, by investigating the typical residual scatter present in plots of
log $\epsilon$(\ion{Fe}{1}) versus log(EW/$\lambda$), and by examining the 
typical line--to--line scatter in the derived log $\epsilon$(\ion{Fe}{1}) and 
log $\epsilon$(\ion{Fe}{2}) abundances.  In order to estimate stellar 
parameters from photometry, we have to assume a cluster distance and reddening.
Ortolani et al. (2000) and Alves et al. (2001) estimate distances of 11.2 kpc 
and 10.7 kpc, respectively, and for this work we have adopted a distance of 
10.7 kpc.  For the reddening, we note that while the differential reddening 
toward NGC 5986 is relatively low (Alves et al. 2001), the absolute reddening 
value is moderately high with literature estimates ranging from approximately 
E(B--V) = 0.22 mag. to 0.36 mag. (Zinn 1980; Bica \& Pastoriza 1983; Rosenberg 
et al. 2000; Alves et al. 2001; Recio--Blanco et al. 2005).  We initially 
assumed E(B--V) = 0.29 mag., which is the mean of the full literature range, 
but subsequent tests showed that the best agreement between the spectroscopic 
and photometric T$_{\rm eff}$ values occurred with E(B--V) = 0.33 mag.  

Using the V magnitudes from Alves et al. (2001) and the 2MASS J and K$_{\rm S}$ 
magnitudes, we followed the photometric transformation procedure outlined in 
Johnson et al. (2005; see their Section 3.1) to obtain photometric temperatures
for all NGC 5986 members.  Assuming E(B--V) = 0.33 mag., we found the 
star--to--star dispersion to be 72 K.  Similarly, assuming a distance of 10.7
kpc and a typical stellar mass of 0.8 M$_{\odot}$\footnote{Although we are 
assuming the same mass for all stars in the photometric surface gravity
calculation, Figure \ref{f1} shows that several stars are likely on the AGB and
may have masses of $\sim$0.6 M$_{\odot}$.  However, the photometric gravity is
only sensitive to log(M/M$_{\odot}$) so the difference in log(g) is only
$\sim$0.10 dex.}, we found the average difference in photometric and 
spectroscopic gravities to be 0.08 dex in log(g), with a dispersion of 0.22 
dex.  Therefore, we have adopted $\Delta$T$_{\rm eff}$ and $\Delta$log(g) of 
75 K and 0.20 dex, respectively, as typical uncertainty values.  For the model 
metallicity uncertainties, we found the average line--to--line scatter in 
log $\epsilon$(\ion{Fe}{1}) and log $\epsilon$(\ion{Fe}{2}) to be 0.10 dex 
($\sigma$ = 0.02 dex) and 0.11 dex ($\sigma$ = 0.04 dex), respectively.  We 
have adopted 0.10 dex as the typical uncertainty in a star's model atmosphere 
metallicity.  Finally, an examination of the line--to--line scatter in plots 
of log $\epsilon$(X) versus log(EW/$\lambda$) suggests that the typical 
microturbulence uncertainty is approximately 0.10 km s$^{\rm -1}$.

In order to estimate the impact of model atmosphere uncertainties on the 
abundance measurements, we redetermined the abundances of each element after
changing T$_{\rm eff}$ $\pm$ 75 K, log(g) $\pm$ 0.20 dex, [Fe/H]\footnote{Note 
that the use of $\alpha$--enhanced model atmospheres largely compensates for 
differences between the iron abundance ([Fe/H]) and overall metallicity 
([M/H]).} $\pm$ 0.10 dex, and $\xi$$_{\rm mic.}$ $\pm$ 0.10 km s$^{\rm -1}$.
Abundance uncertainty terms were calculated for each element by individually
varying the model atmosphere parameters while holding the other values
fixed.  The abundance uncertainties due to varying each model atmosphere 
parameter were then added in quadrature, along with the line--to--line 
measurement uncertainties, and are provided in Tables 3--4.

Finally, we note that an investigation into trends of [X/Fe] versus 
T$_{\rm eff}$ and [Fe/H] revealed that minor trends may exist between [Ca/Fe]
and T$_{\rm eff}$ and [Eu/Fe] and [Fe/H].  For Ca, we find that stars with
T$_{\rm eff}$ $<$ 4400 K have $\langle$[Ca/Fe]$\rangle$ = $+$0.29 dex 
($\sigma$ = 0.03 dex) while those with T$_{\rm eff}$ $>$ 4400 K have 
$\langle$[Ca/Fe]$\rangle$ = $+$0.23 dex ($\sigma$ = 0.03 dex).  However, the
star--to--star dispersion in [Ca/Fe] is only 0.04 dex for the entire sample, 
and we did not find any trends between T$_{\rm eff}$ and 
log $\epsilon$(\ion{Fe}{1}), log $\epsilon$(\ion{Fe}{2}), nor 
log $\epsilon$(\ion{Ca}{1}).  Combined with the paucity of similar trends 
between T$_{\rm eff}$ and any other [X/Fe] ratios, we believe that the 
mild correlation between T$_{\rm eff}$ and [Ca/Fe] is insignificant.

The correlation between [Eu/Fe] and [Fe/H] is more troubling as stars with
[Fe/H] $<$ --1.55 have $\langle$[Eu/Fe]$\rangle$ = $+$0.69 dex 
($\sigma$ = 0.04 dex) and those with [Fe/H] $>$ --1.55 have
$\langle$[Eu/Fe]$\rangle$ = $+$0.81 dex ($\sigma$ = 0.07 dex).  Since the 
other elements examined here do not exhibit similarly strong correlations 
with [Fe/H], we do not have a clear explanation for the behavior of [Eu/Fe].
For example, a simple explanation such as improperly accounting for a blend
of the \ion{Si}{1} line at 6437.71 \AA\ and the \ion{Eu}{2} line at 6437.64 
\AA\ is unlikely because [Si/Fe] is not correlated with [Fe/H] nor [Eu/Fe].
In any case, we caution the reader that the observed star--to--star variation
of $\sim$0.1 dex presented here for [Eu/Fe] may be an overestimate.

\section{RESULTS AND DISCUSSION}

\subsection{Metallicity Distribution Function}

As mentioned in Section 1, the analysis of two AGB/post--AGB stars by
Jasniewicz et al. (2004), which found a cluster metallicity of [Fe/H] $\sim$ 
--1.65, represents the only detailed chemical composition measurement of 
individual stars in NGC 5986.  However, previous and subsequent photometric 
analyses have found in agreement that the cluster has a mean metallicity of 
[Fe/H] $\sim$ --1.60 (Bica \& Pastoriza 1983; Geisler et al. 1997; Ortolani
et al. 2000; Dotter et al. 2010).  In this work, we measured [Fe/H] for 25 RGB
and AGB stars and derived a similar mean metallicity of 
$\langle$[Fe/H]$\rangle$ = --1.54 dex ($\sigma$ = 0.08 dex).

\begin{figure}
\epsscale{1.00}
\plotone{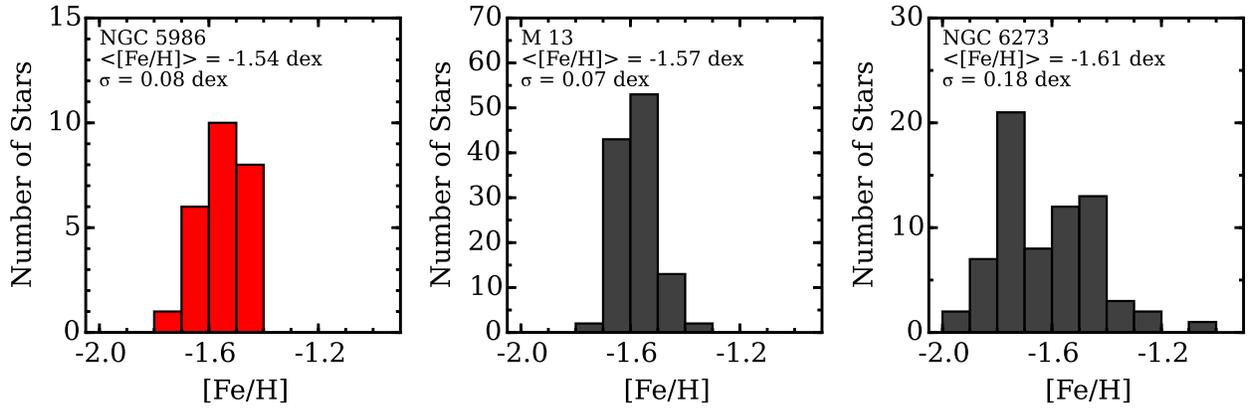}
\caption{The three panels compare the metallicity distribution functions of
NGC 5986 (left panel), M 13 (Johnson \& Pilachowski 2012; middle panel), and
NGC 6273 (Johnson et al. 2015a, 2017; right panel), which were derived using 
similar methods and line lists and also exhibit similar mean metallicities.  
The [Fe/H] abundance spread for NGC 5986 is consistent with other monometallic 
clusters, such as M 13.  The data for all three clusters are sampled into
0.10 dex bins.}
\label{f3}
\end{figure}

Interestingly, although NGC 5986 has a mean metallicity, present--day mass,
and horizontal branch morphology that is similar to several iron--complex
clusters, its comparatively small [Fe/H] dispersion likely precludes the 
cluster from being a member of the iron--complex class.  For example, 
Figure \ref{f3} compares the [Fe/H] distributions of the monometallic cluster
M 13, the iron--complex cluster NGC 6273, and NGC 5986.  All three clusters
exhibit extended blue horizontal branches, are relatively massive, and have
comparable mean [Fe/H] values, but Figure \ref{f3} shows that NGC 5986 lacks 
the broad [Fe/H] spread that is a defining characteristic of iron--complex 
clusters.  Instead, NGC 5986 appears to be a more typical monometallic cluster,
similar to M 13.

\subsection{Basic Chemical Composition Results}

\begin{figure}
\epsscale{0.75}
\plotone{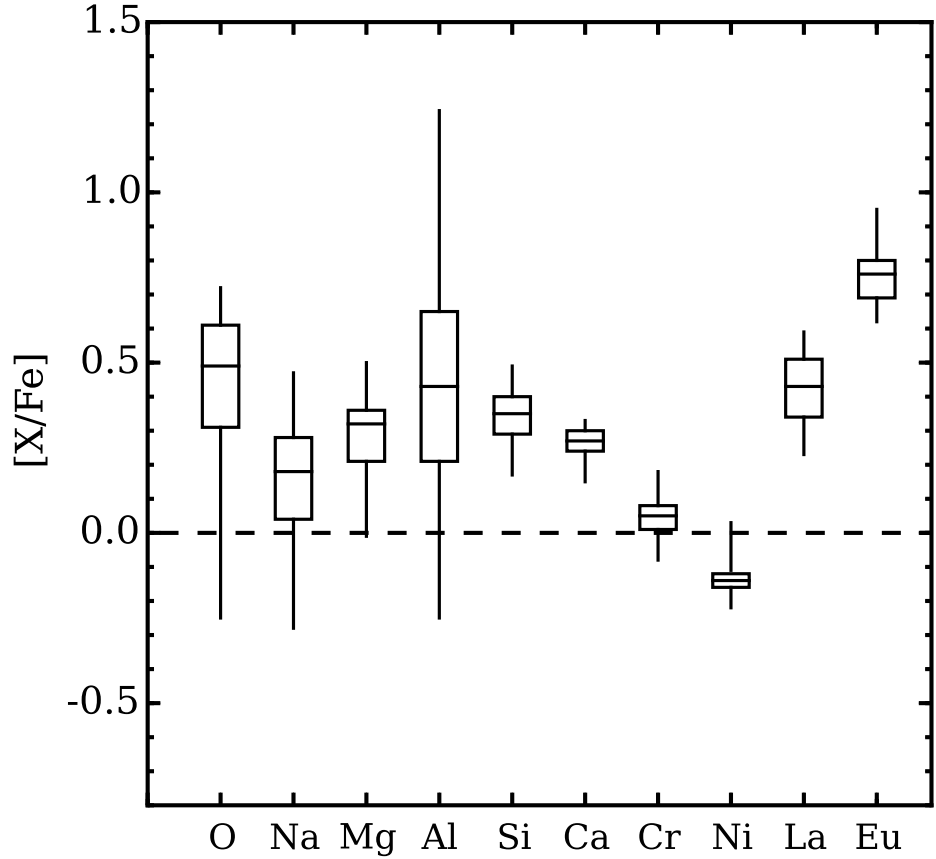}
\caption{This box plot compares the [X/Fe] distributions of all elements 
analyzed here for the NGC 5986 radial velocity member stars.  For each element,
the bottom, middle, and top horizontal lines indicate the first, second 
(median), and third quartile values, respectively.  The extended vertical
lines indicate the minimum and maximum [X/Fe] abundances.  The black dashed
line illustrates the Solar abundance ratios.}
\label{f4}
\end{figure}

A summary of the chemical abundances found in NGC 5986 is provided as a box 
plot in Figure \ref{f4}.  Similar to other old globular clusters (e.g., see 
reviews by Kraft 1994 and Gratton et al. 2004), the light elements O, Na, Mg, 
and Al exhibit the largest star--to--star abundance variations (see also
Section 5.3).  Additionally, the heavy $\alpha$--elements are enhanced with 
$\langle$[Si/Fe]$\rangle$ = $+$0.34 dex ($\sigma$ = 0.08 dex) and 
$\langle$[Ca/Fe]$\rangle$ = $+$0.27 dex ($\sigma$ = 0.04 dex) while the 
Fe--peak elements Cr and Ni have approximately Solar ratios with 
$\langle$[Cr/Fe]$\rangle$ = $+$0.04 dex ($\sigma$ = 0.06 dex) and 
$\langle$[Ni/Fe]$\rangle$ = --0.12 dex ($\sigma$ = 0.06 dex).  Interestingly, 
the two neutron--capture elements La and Eu are both enhanced with 
$\langle$[La/Fe]$\rangle$ = $+$0.42 dex ($\sigma$ = 0.11 dex) and 
$\langle$[Eu/Fe]$\rangle$ = $+$0.76 dex ($\sigma$ = 0.08 dex).  We detect a 
mild correlation between [La/Fe] and [Eu/Fe], and the full [X/Fe] range 
exhibited by each element is $\sim$0.35 dex.  Following Roederer (2011), we
consider NGC 5986 to be a borderline case that may possess a small intrinsic 
heavy element dispersion\footnote{As discussed in Section 4.2, we detected a 
possible correlation between [Fe/H] and [Eu/Fe] that may be spurious and 
caution the reader that the cluster's true [Eu/Fe] dispersion may be smaller
than the 0.08 dex value cited here.}.  However, the available data seem to 
rule out that NGC 5986 is similar to more extreme cases, such as M 15, where 
the full range in [La/Fe] and [Eu/Fe] can span $>$ 0.6 dex (e.g., Sneden et 
al. 1997; Sobeck et al. 2011).  

Although the present work represents the only large sample chemical abundance
analysis of cluster RGB stars, we note that Jasniewicz et al. (2004) measured 
O, Na, Mg, Si, Ca, Cr, Ni, La, and Eu abundances for two highly evolved 
AGB/post--AGB stars.  In general, both studies agree that NGC 5986 stars have 
[$\alpha$/Fe] $\sim$ $+$0.30 dex, approximately Solar [X/Fe] ratios for the 
Fe--peak elements, and enhanced [La/Fe] and [Eu/Fe] ratios.  However, we find 
an average $\langle$[Na/Fe]$\rangle$ = $+$0.15 dex, which is significantly 
lower than the [Na/Fe] = $+$0.70--1.00 dex ratios measured by Jasniewicz et 
al. (2004).  The heavy neutron--capture element abundance pattern
exhibited by the potential post--AGB star NGC 5986 ID7 in Jasniewicz et al.
(2004) follows the same distribution as the RGB/AGB stars measured here.  The
low [La/Eu] ratios found in both studies suggest that NGC 5986 did not 
experience significant s--process enrichment, and therefore the enhanced 
[Zr/Fe], [La/Fe], [Ce/Fe], [Sm/Fe], and [Eu/Fe] abundances of ID7 may not 
necessarily reflect additional processing and/or mixing via third dredge--up.

\begin{figure}
\epsscale{1.00}
\plotone{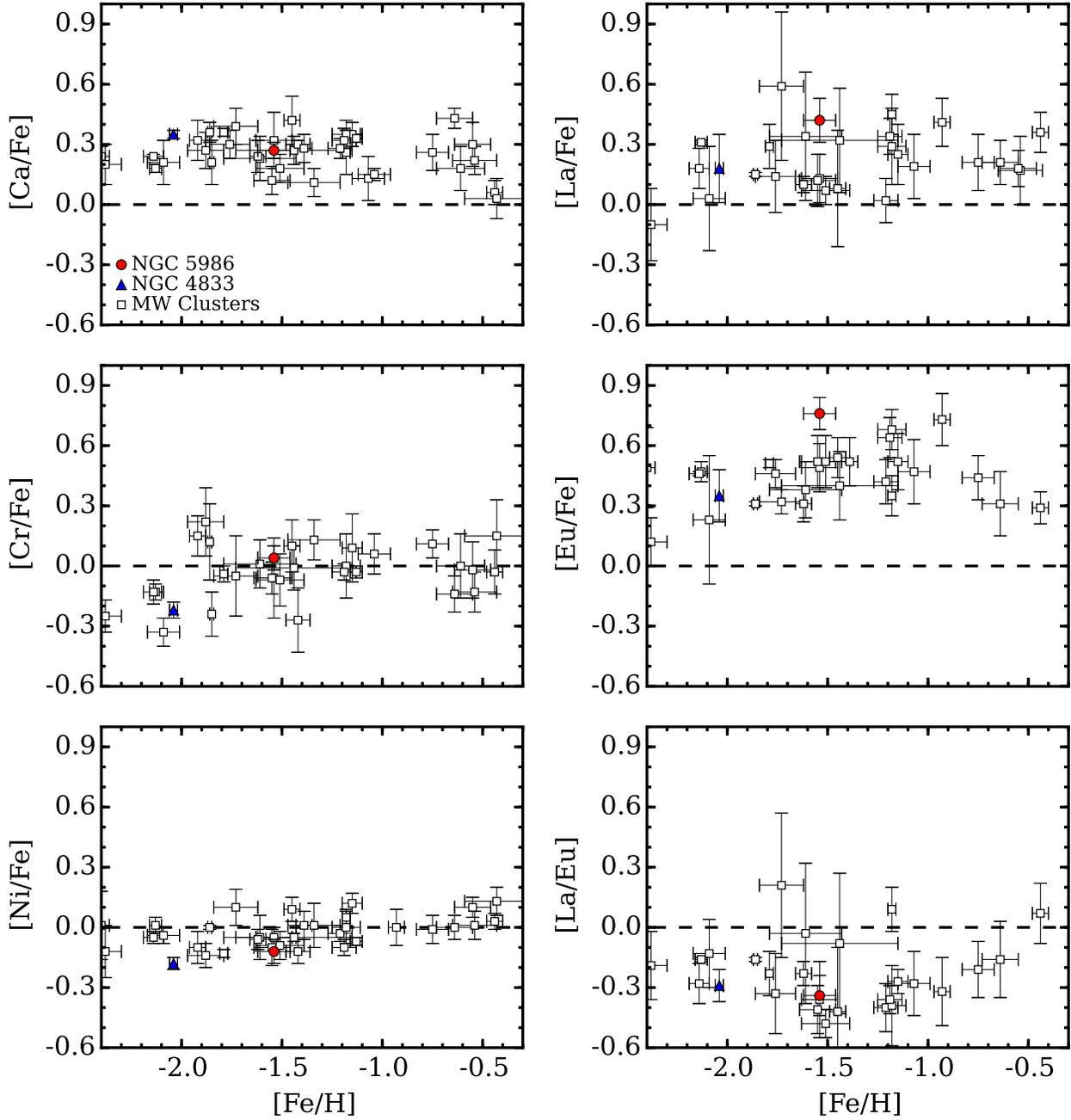}
\caption{The average [Ca/Fe], [Cr/Fe], [Ni/Fe], [La/Fe], [Eu/Fe], and [La/Eu]
abundances of NGC 5986 (filled red circles) and NGC 4833 (filled blue
triangles) are compared to those of several Galactic globular clusters (open
boxes) of different [Fe/H].  In all panels, the symbols indicate the
cluster average values and the error bars show the star--to--star dispersions.
The literature sources for each globular cluster are provided in Table 5.}
\label{f5}
\end{figure}

In Figure \ref{f5}, we compare the abundance pattern of NGC 5986 against 
several Milky Way globular clusters spanning a wide range in [Fe/H], and find
that NGC 5986 has a composition that is nearly identical to other similar 
metallicity clusters.  In fact, only the cluster's [La/Fe] and [Eu/Fe] 
abundances deviate from the typical Galactic trend with NGC 5986 stars 
exhibiting higher ratios; however, the cluster's heavy element composition may 
bear some resemblance to that of M 107 (O'Connell et al. 2011).  The 
combination of enhanced [La/Fe] and [Eu/Fe] found in NGC 5986 also nearly 
matches the pattern found by Cavallo et al. (2004) for the similar metallicity 
cluster M 80, but a recent analysis by Carretta et al. (2015) revised M 80's 
mean [La/Fe] and [Eu/Fe] abundances downward by $\sim$0.3 dex.  

\begin{figure}
\epsscale{1.00}
\plotone{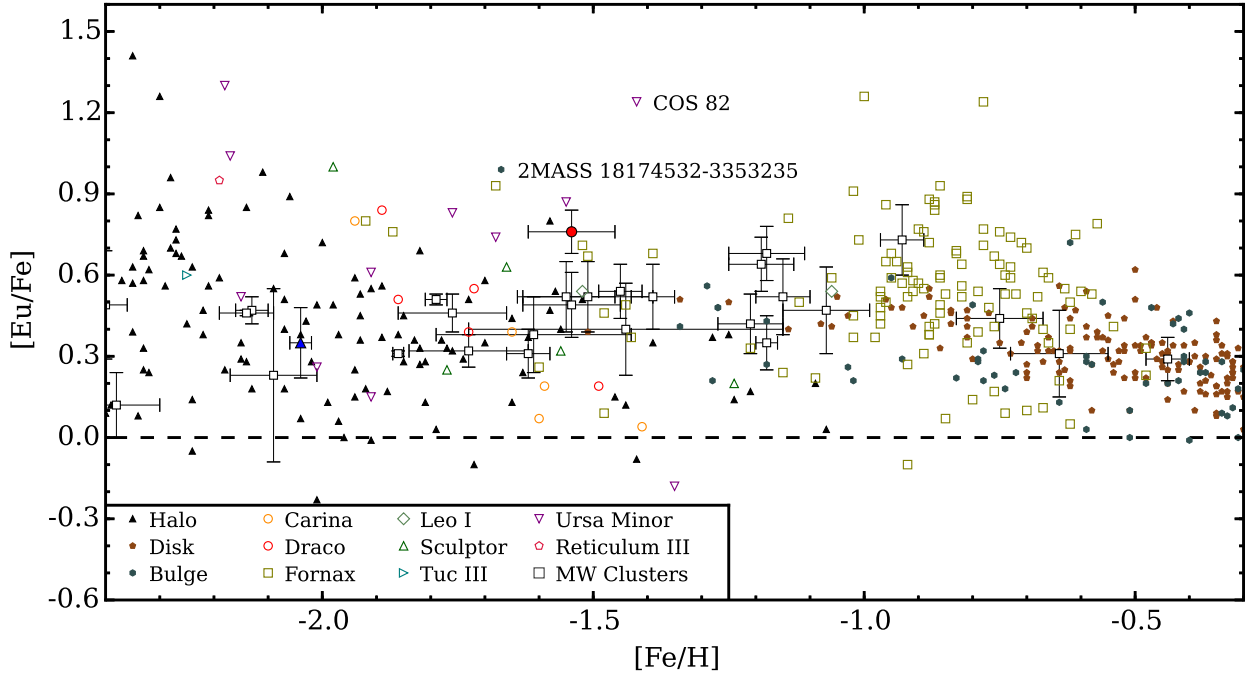}
\caption{The mean [Eu/Fe] abundances of NGC 5986 (filled red circle), NGC 4833
(filled blue triangle), and several Galactic globular clusters (open boxes with
error bars) are plotted as a function of [Fe/H].  The globular clusters are 
compared against similar composition measurements in the Galactic halo, disk, 
and bulge along with several classical and ultra--faint Local Group dwarf 
galaxies.  The literature data for the halo, thin/thick disk, and bulge are 
compiled from Barklem et al. (2005), McWilliam et al. (2010), Johnson et al. 
(2012,2013), Roederer et al. (2014), Battistini \& Bensby (2016), and Van der 
Swaelmen et al. (2016).  The dwarf galaxy data are from: Shetrone et al. (2003)
for Carina, Shetrone et al. (2001) and Cohen \& Huang (2009) for Draco, 
Shetrone et al. (2003), Letarte et al. (2010), and Lemasle et al. (2014) for 
Fornax, Shetrone et al. (2003) for Leo I and Sculptor, Hansen et al. (2017) 
for Tucana III, Shetrone et al. (2001), Aoki et al. (2007), and Cohen \& Huang 
(2010) for Ursa Minor, and Ji et al. (2016) for Reticulum II.  The globular 
cluster data are the same as in Figure \ref{f5}.  Stars in the Galactic bulge 
and Ursa Minor dwarf galaxy that have [Fe/H] similar to NGC 5986 but [Eu/Fe] 
$\ga$ $+$1 are also identified in the plot.}
\label{f6}
\end{figure}

To place NGC 5986 into context, Figure \ref{f6} extends the [Eu/Fe] panel of 
Figure \ref{f5} to include a comparison of the cluster against individual 
stars in the Galactic halo, thin/thick disk, bulge, and several Local Group
classical and ultra--faint dwarf galaxies.  Although it is not unusual to find
strongly Eu--enhanced stars at [Fe/H] $\ga$ --2 dex, such objects are almost 
exclusively found in dwarf galaxies exhibiting large [Eu/Fe] dispersions.  In
contrast, NGC 5986 exhibits a mean [Eu/Fe] abundance that is comparable to 
some of the most Eu--enhanced stars in galaxies such as Fornax, Carina, Ursa 
Minor, and Draco, but the cluster's [Eu/Fe] dispersion is at least 1.5--2 times
smaller.  The data suggest NGC 5986 was enriched by a high--yield r--process 
event (or events) and that the enriched gas was able to be rapidly mixed 
within the cluster.

\begin{figure}
\epsscale{1.00}
\plotone{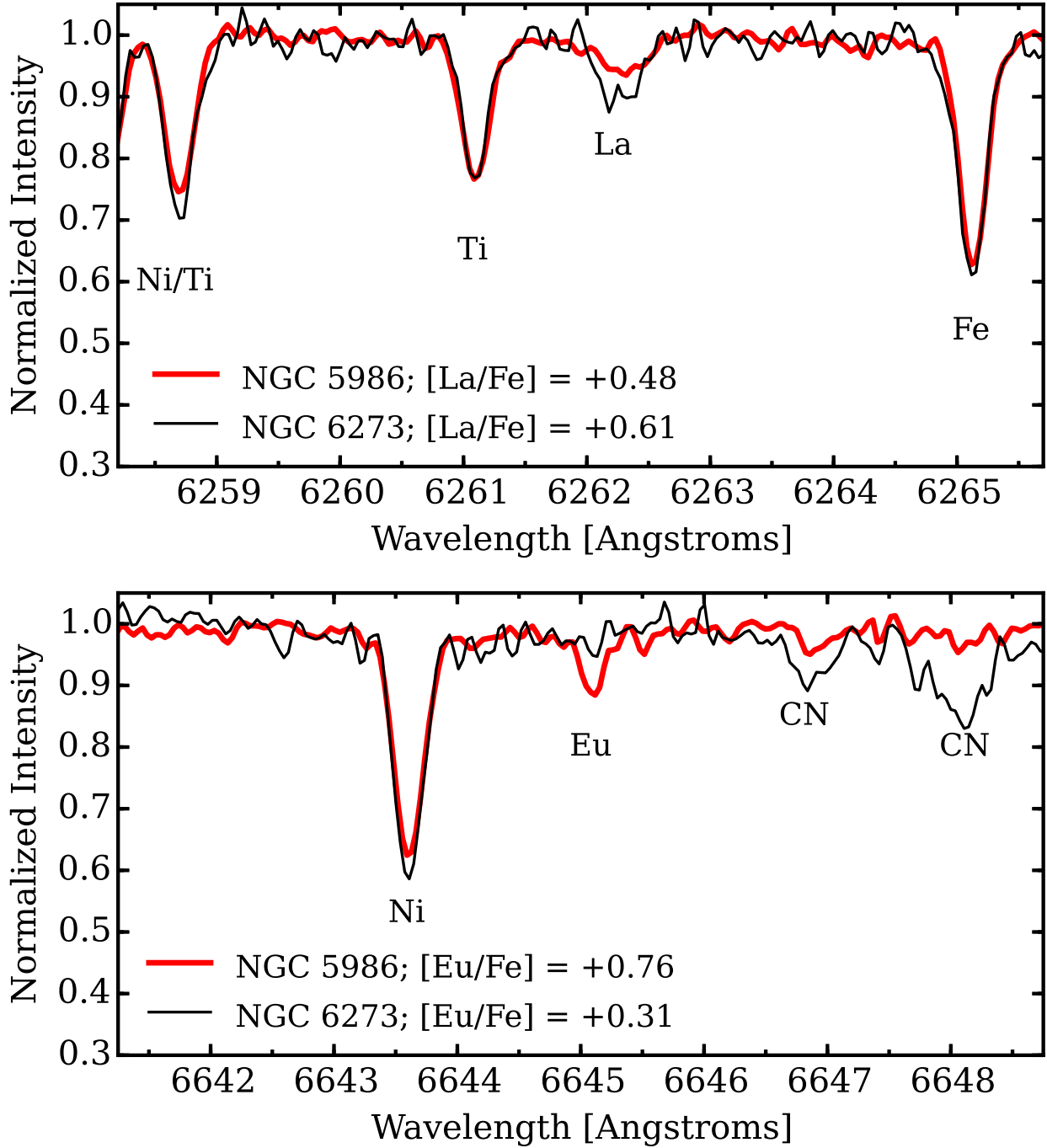}
\caption{The top and bottom panels compare spectra of the NGC 5986 star
2MASS 15460476--3749186 (this work) and the NGC 6273 star 
2MASS 17024412--2616495 (Johnson et al. 2017) for regions near the 6262 \AA\ 
\ion{La}{2} and 6645 \AA\ \ion{Eu}{2} lines.  The two stars have similar 
temperatures, gravities, metallicities, and [La/Fe] abundances but different 
[Eu/Fe] abundances.  Note that the NGC 6273 star exhibits stronger CN lines 
because it has a different CNO composition.}
\label{f7}
\end{figure}

Interestingly, the high [La/Fe] abundances of stars in NGC 5986 are similar to 
those found in the more metal--rich populations of the iron--complex clusters 
$\omega$ Cen (e.g., Norris \& Da Costa 1995; Smith et al. 2000; Johnson \& 
Pilachowski 2010; Marino et al. 2011b), NGC 5286 (Marino et al. 2015), M 22 
(Marino et al. 2009, 2011a), M 2 (Yong et al. 2014a), NGC 1851 (Carretta et al. 
2011), and NGC 6273 (Johnson et al. 2015a, 2017).  The mean metallicity of NGC 
5986 is comparable to the typical [Fe/H] values exhibited by the 
Fe/s--process enhanced populations in several iron--complex clusters as well.  
For example, Figure \ref{f7} compares the spectrum of a star in NGC 5986 with
the spectrum of an Fe/s--process enhanced star in NGC 6273 of similar 
temperature, metallicity, and gravity, and shows that both objects have 
comparable [Fe/H] and [La/Fe] abundances.  However, Figure \ref{f7} 
also shows that the 6645 \AA\ \ion{Eu}{2} line in the NGC 5986 star is 
considerably stronger than in the NGC 6273 star.  The cluster's high [Eu/Fe]
abundances and mean $\langle$[La/Eu]$\rangle$ = --0.34 dex ($\sigma$ = 0.10 dex)
therefore follow the same r--process dominated pattern exhibited by most
monometallic clusters and precludes NGC 5986 from being an iron--complex 
cluster.

Figure \ref{f5} also compares the $\alpha$, Fe--peak, and neutron--capture 
element abundances of NGC 5986 and NGC 4833.  Although NGC 4833 is 
significantly more metal--poor than NGC 5986 at [Fe/H] $\approx$ --2.15
(Carretta et al. 2014; Roederer \& Thompson 2015), Casetti--Dinescu et al.
(2007) found the two clusters to exhibit similar orbital properties and 
suggested that the clusters may share a common origin.  With the present data,
it is difficult to assess whether NGC 5986 and NGC 4833 are both chemically
and dynamically linked.  For example, Figure \ref{f5} shows that the two 
clusters share similar mean [Ca/Fe], [Ni/Fe], and [La/Eu] abundances, and we 
can also note that both clusters may exhibit larger than average [Ca/Mg]
dispersions (see also Section 5.3 here and Figure 15 of Carretta et al. 2014).
In contrast, the clusters appear to have very different mean [Cr/Fe], [La/Fe],
and [Eu/Fe] abundances.  Therefore, NGC 5986 and NGC 4833 may exhibit larger
cluster--to--cluster heavy element abundance variations than are observed among
similar metallicity globular clusters associated with the Sagittarius
system (e.g., Mottini et al. 2008; Sbordone et al. 2015).  However, if NGC 
5986 and NGC 4833 do originate from a common system, such as a dwarf galaxy, 
then the progenitor object may have followed a chemical enrichment path that 
differs from Sagittarius.

\subsection{Light Element Abundance Variations}

As mentioned in Section 5.2, the light elements O, Na, Mg, and Al exhibit
significant star--to--star abundance variations with $\Delta$[X/Fe] ranging
from 0.51 dex for [Mg/Fe] to 1.49 dex for [Al/Fe].  Additionally, 
Figure \ref{f8} shows that NGC 5986 exhibits a clear O--Na anti--correlation 
along with a strong Na--Al correlation.  The O--Na and Na--Al relations are 
common features found in nearly all clusters with [Fe/H] $\la$ --1, and the
presence of these (anti--)correlations in stars at all evolutionary states
(see Section 1) indicates that pollution, rather than \emph{in situ} mixing,
is the dominant mechanism driving the light element abundance variations.  In
this sense, the simultaneous signature of O--depletions with enhancements in Na
and Al indicate that the gas from which the second generation (O/Mg--poor; 
Na/Al--rich) stars formed was processed at temperatures of at least $\sim$45 MK
(e.g., Prantzos et al. 2007; their Figure 2).  However, a variety of pollution 
sources, such as intermediate mass AGB stars (e.g., Ventura \& D'Antona 2009; 
Doherty et al. 2014), rapidly rotating massive main--sequence stars (e.g., 
Decressin et al. 2007), massive interacting binary stars (de Mink et al. 2009),
and supermassive stars (Denissenkov \& Hartwick 2014), are capable of reaching
these temperatures.  Therefore, an examination of the interplay between Mg, Al,
and Si, which are more sensitive to higher burning temperatures, can help
shed light on which pollution mechanism(s) may have been active in NGC 5986 and 
other clusters.

The bottom panels of Figure \ref{f8} show that NGC 5986 exhibits a clear 
Mg--Al anti--correlation and Al--Si correlation.  Unlike the more ubiquitous
O--Na and Na--Al relations, the Mg--Al and Al--Si (anti--)correlations are not
found in all clusters.  Instead, these chemical properties seem to be common
only among massive and/or metal--poor clusters with extended blue horizontal 
branches, such as: NGC 2808 (Carretta 2014, 2015), NGC 6752 (Yong et al. 2005; 
Carretta et al. 2012a), M 15 (Sneden et al. 1997; Carretta et al. 2009b); M 13 
(Sneden et al. 2004; Cohen \& Mel{\'e}ndez 2005), NGC 6273 (Johnson et al. 
2015a, 2017), $\omega$ Cen (Norris \& Da Costa 1995; Smith et al. 2000; Da 
Costa et al. 2013), M 54 (Carretta et al. 2010b), NGC 1851 (Carretta et al. 
2012b), NGC 4833 (Carretta et al. 2014; Roederer \& Thompson 2015), M 92 
(M{\'e}sz{\'a}ros et al. 2015; Ventura et al. 2016), and NGC 6093 (Carretta
et al. 2015).  In NGC 5986 and similar clusters, the significant depletion of 
\iso{24}{Mg} requires temperatures $\ga$65--70 MK (e.g., Langer et al. 1997; 
Arnould et al. 1999; Prantzos et al. 2007).  At these temperatures the 
\iso{27}{Al}(p,$\gamma$)\iso{28}{Si} reaction rate exceeds 
that of the \iso{27}{Al}(p,$\alpha$)\iso{24}{Mg} reaction (e.g., see Arnould et
al. 1999; their Figure 8), and leakage from the Mg--Al cycle can produce small 
increases ($\sim$0.1 dex) in [Si/Fe].  Therefore, despite significant 
uncertainties in several reaction rates (e.g., Izzard et al. 2007), the 
combined abundance patterns of O, Na, Mg, Al, and Si support the idea that the 
gas from which the second generation stars in NGC 5986 formed likely 
experienced temperatures $\ga$65--70 MK.  Following D'Antona et al. (2016), we 
can conclude that only the AGB and/or supermassive star pollution scenarios 
mentioned above likely remain viable to explain the light element patterns of 
NGC 5986; however, these scenarios still face substantial challenges in 
explaining all of the observed abundance patterns (e.g., Renzini et al. 2015).

In clusters such as NGC 2419 (Cohen \& Kirby 2012; Mucciarelli et al. 2012) and
NGC 2808 (Carretta 2015; Mucciarelli et al. 2015c), the O, Na, Mg, Al, and Si
abundance (anti--)correlations are accompanied by similar relations involving
elements as heavy as K, Ca, and Sc.  For these cases, Ventura et al. (2012) 
noted that the abundance patterns may be explained if proton--capture reactions
operated in an environment where the burning temperatures exceeded 
$\sim$100 MK.  Although we did not measure K and Sc abundances for 
NGC 5986, we note that Carretta et al. (2013b) and Carretta (2015) have shown
that the [Ca/Mg] spread may be a reliable indicator for finding clusters that
experienced extreme high temperature processing.  For NGC 5986, we find that
[Ca/Mg] ranges from --0.17 dex to $+$0.33 dex, which is larger than many
clusters but less extreme than NGC 2419 and NGC 2808 (e.g., see Carretta 2015; 
their Figure 14).  Therefore, we conclude that the gas from which the second 
generation stars in NGC 5986 formed did not experience significant processing
at temperatures $>$100 MK and was instead limited to $\sim$70--100 MK.

Regardless of the exact pollution source(s), multiple studies agree that 
dilution/mixing between the ejected material and pre--existing gas is likely
required to explain the observed abundance patterns (e.g., D'Antona \& Ventura
2007; Decressin et al. 2007; D'Ercole et al. 2008; Denissenkov \& Hartwick 
2014; D'Antona et al. 2016; but see also Bastian et al. 2015).  In particular,
the shapes of the various light element anti--correlations in many globular 
clusters closely resemble simple dilution curves.  However, attempts to fit
the observed data with simple dilution models have largely failed (Carretta
et al. 2012a; Bastian et al. 2015; Carretta 2014; Villanova et al. 2017).  In 
this light, Figure \ref{f9} shows the O--Na, Na--Mg, and Mg--Al 
anti--correlations for NGC 5986 along with dilution curves that represent the 
expected distributions from mixing first generation compositions with the most 
extreme second generation compositions.  Although the O--Na and Na--Mg 
anti--correlations are nearly aligned with the expected distributions, neither 
presents an exact match.  Compared to the dilution curves, the typical [Na/Fe] 
abundances of many intermediate composition stars are systematically too high 
for a given [O/Fe] or [Mg/Fe] value.  Furthermore, the Mg--Al dilution curve is
a poor fit to the data.  Similar to the cases of NGC 6752 and NGC 2808 
(Carretta et al. 2012a; Carretta 2015), we conclude that the light element 
distributions in NGC 5986 are not well--described by a simple dilution scenario
and that more than one significant pollution source must have been present. 

\subsection{Discrete Populations}

\begin{figure}
\epsscale{1.00}
\plotone{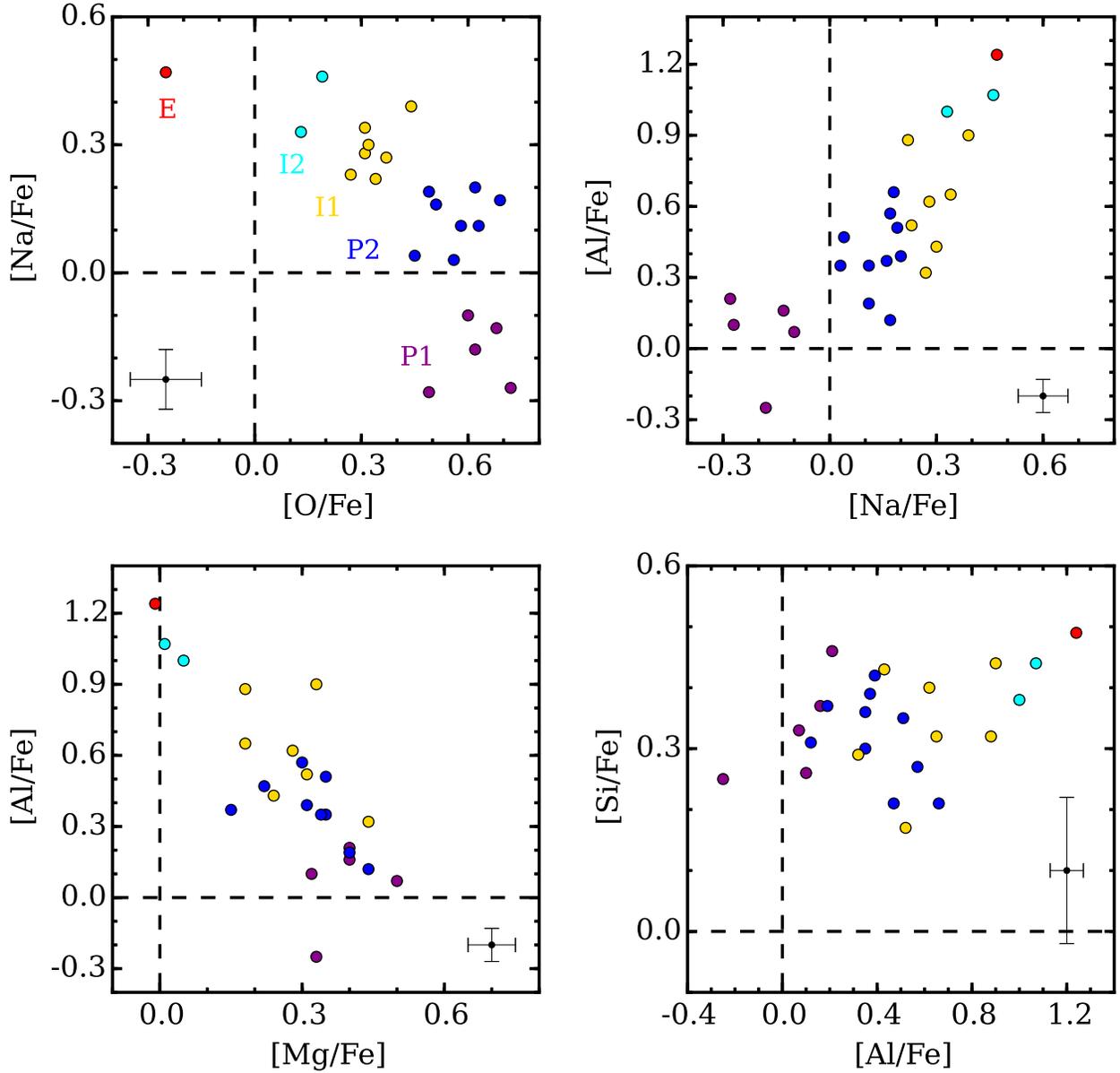}
\caption{[O/Fe], [Na/Fe], [Mg/Fe], [Al/Fe], and [Si/Fe] are plotted against 
each other to illustrate the O--Na/Mg--Al anti--correlations and Na--Al/Al--Si
correlations present in NGC 5986.  The [Na/Fe] versus [O/Fe] plot in particular
suggests that $\sim$5 distinct populations with different light element 
chemistry may exist.  We have labeled and color--coded the ``primordial" 
(\emph{P1}; \emph{P2}), ``intermediate" (\emph{I1}; \emph{I2}), and ``extreme"
(\emph{E}) groups based on the nomenclature used in Carretta (2015) for 
NGC 2808.  The existence of an Al--Si correlation suggests that the gas from 
which the more Al--rich stars formed was processed at temperatures 
$\ga$ 65 MK.}
\label{f8}
\end{figure}

\begin{figure}
\epsscale{1.00}
\plotone{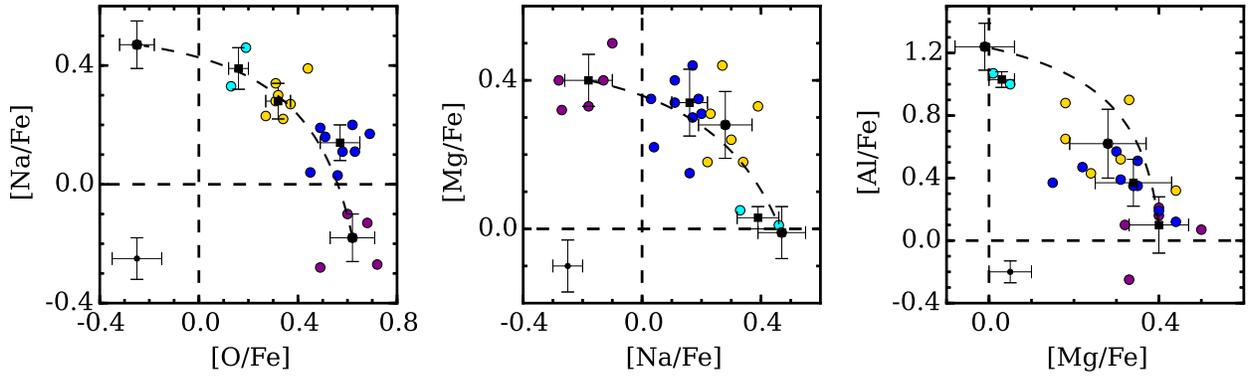}
\caption{The three panels illustrate the shape and extent of the O--Na (left), 
Na--Mg (middle), and Mg--Al (left) anti--correlations present in NGC 5986.  
The filled black boxes indicate the median [X/Fe] ratios for each of the 
populations identified in Figure \ref{f8}, and the error bars represent
the approximate [X/Fe] dispersion within each population.  The dashed black
lines are dilution curves with the end points anchored at the median [X/Fe] 
values of the \emph{P1} and \emph{E} populations.  The remaining symbols are
the same as those in Figure \ref{f8}.}
\label{f9}
\end{figure}

In addition to ruling out a simple dilution model, Figures \ref{f8}--\ref{f9} 
indicate that the light element abundances in NGC 5986 may separate into
discrete groups rather than follow a continuous distribution.  Using the
O--Na panel of Figure \ref{f8} and following the nomenclature of Carretta
(2015) for NGC 2808, we have identified five possible unique populations in
NGC 5986 that are labeled as \emph{P1}, \emph{P2}, \emph{I1}, \emph{I2}, and
\emph{E}.  In this scheme, the ``primordial" \emph{P1} and \emph{P2} groups 
have higher [O/Fe] and [Mg/Fe] and lower [Na/Fe], [Al/Fe], and [Si/Fe] than
the ``intermediate" \emph{I1} and \emph{I2} groups while the ``extreme" 
\emph{E} population (1 star) exhibits the lowest [O/Fe] and [Mg/Fe] and highest
[Na/Fe], [Al/Fe], and [Si/Fe] abundances.  Assuming each of the identified
groups is real, the \emph{P1}, \emph{P2}, \emph{I1}, \emph{I2}, and \emph{E}
populations constitute approximately 20$\%$, 40$\%$, 28$\%$, 8$\%$, and 4$\%$
of our sample, respectively.  Interestingly, the combined 60$\%$ fraction of 
first generation composition stars is similar to NGC 2808 (Carretta 2015), 
which is another cluster known to host at least five populations.  Both 
NGC 2808 and NGC 5986 seem to host anomalously large fractions of first 
generation stars.

Although the various populations are relatively well--separated in the O--Na 
panels of Figures \ref{f8}--\ref{f9}, the composition boundaries distinguishing
all five populations are less clear for other elements.  Specifically, the 
Na--Al, Na--Mg, and Mg--Al plots provide strong evidence that at least three 
discrete populations exist (\emph{P1}, \emph{P2 $+$ I1}, and \emph{I2 $+$ E}), 
but the similar abundance patterns between especially the \emph{P2} and 
\emph{I1} groups make additional separations more ambiguous.  In order to 
more quantitatively assess the likelihood that at least 4--5, rather than 3,
unique populations exist, we utilized the Welch's \emph{t}--test to examine
differences in the mean light element [X/Fe], [O/Na], [Na/Mg], [Mg/Al], and 
[Al/Si] distributions of the \emph{P1}, \emph{P2}, \emph{I1}, and \emph{I2} 
groups.  The \emph{E} population only constitutes one star and is omitted from
this analysis; however, since this star has the lowest [O/Fe] and [Mg/Fe] 
and highest [Na/Fe], [Al/Fe], and [Si/Fe] abundances, we consider the 
\emph{E} population to be separate from the \emph{I2} stars.

A summary of the Welch \emph{t}--test results is provided in Table 6 and 
generally reinforces the original hypothesis that up to five populations may
exist.  If we adopt the common convention that a \emph{p}--value $<$ 0.05
indicates sufficient evidence exists to reject the null hypothesis that two
element/population pairs exhibit the same mean composition, then nearly all 
of the comparisons in Table 6 support $>$ 3 populations being present in
NGC 5986.  A comparison between the \emph{P2} and \emph{I1} groups indicates 
that the two populations may share similar mean [Mg/Fe] and [Si/Fe] abundances,
but the \emph{p}--values for [O/Fe], [Na/Fe], [Al/Fe], [O/Na], and [Na/Mg] are 
below 0.05.  Even though the total sample size is only 25 stars, the 
\emph{P2} and \emph{I1} groups contain 10 and 7 stars, respectively, and are
therefore less sensitive to sampling effects than the other populations.  We
conclude that the \emph{P2} and \emph{I1} groups are likely distinct 
populations.

\begin{figure}
\epsscale{1.00}
\plotone{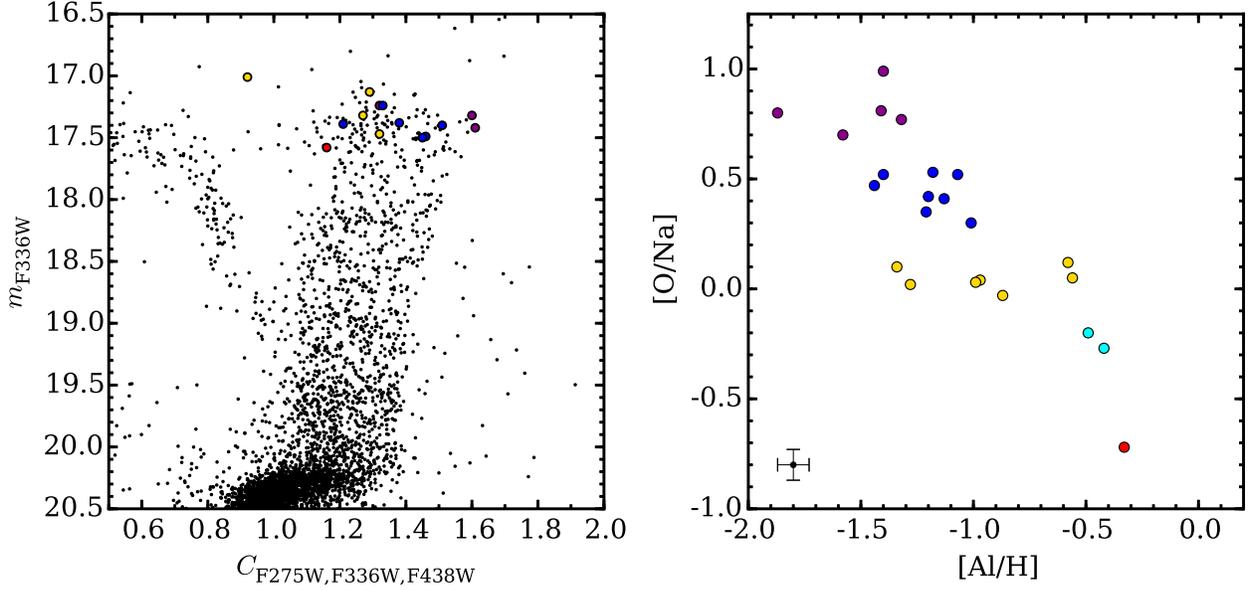}
\caption{\emph{Left:} A \emph{m}$_{\rm F336W}$ versus 
\emph{C}$_{\rm F275W,F336W,F438W}$ pseudo--color--magnitude diagram is shown 
for NGC 5986 using data from Piotto et al. (2015) and Soto et al. (2017; small 
black circles).  The large filled circles follow the same color scheme as in 
Figure \ref{f8}, and represent overlapping stars between the present study and 
the \emph{HST} data.  In general, stars with lower [O/Fe] and [Mg/Fe] and 
higher [Na/Fe] and [Al/Fe] exhibit smaller \emph{C}$_{\rm F275W,F336W,F438W}$ 
pseudo--color values, but some scatter is present due to the combined effects 
of high reddening and the mixing of RGB and AGB stars.  Note that the 
\emph{C}$_{\rm F275W,F336W,F438W}$ pseudo--color index is defined as 
\emph{C}$_{\rm F275W,F336W,F438W}$ $\equiv$ 
(\emph{m}$_{F275W}$--\emph{m}$_{F336W}$)--(\emph{m}$_{F336W}$--\emph{m}$_{F438W}$) (e.g., Milone et al. 2013).  \emph{Right:} Similar to Figure 1 of Gratton
et al. (2011), the various stellar populations of NGC 5986 are distinguished
using a combination of the measured [O/Na] and [Al/H] ratios.  The symbols are
the same as those in Figure \ref{f8}.}
\label{f10}
\end{figure}

As noted in Section 1, \emph{HST} UV--optical photometry is efficient at 
separating globular cluster stars with different light element abundances and
provides an alternative method for examining a cluster's light element 
composition.  In Figure \ref{f10}, we overlay the locations of the different
populations identified here on the \emph{m}$_{\rm F336W}$ versus
\emph{C}$_{F275W,F336W,F438W}$ pseudo--color--magnitude diagram using data
from Piotto et al. (2015) and Soto et al. (2017)\footnote{The \emph{HST} data 
are available for download at: http://groups.dfa.unipd.it/ESPG/treasury.php.  
Note that the \emph{C}$_{F275W,F336W,F438W}$ pseudo--color index is defined as:
(\emph{m}$_{F275W}$--\emph{m}$_{F336W}$)--(\emph{m}$_{F336W}$--\emph{m}$_{F438W}$).}.  Although distinct sequences are not easily separated in Figure \ref{f10} 
from the photometry, perhaps as a result of the moderately high cluster 
reddening (see Section 4.2), the different populations identified via 
spectroscopy tend to cluster in distinct regions.  For example, the 
``primordial" stars tend to exhibit the reddest pseudo--colors while the 
``intermediate" and ``extreme" groups are found at bluer pseudo--colors.  Some
population mixing is observed in Figure \ref{f10}, but we suspect that this is
largely driven by a combination of reddening and the mixing of RGB and AGB 
stars in the sample.

Further evidence supporting the existence of discrete populations in NGC 5986
is shown in the right panel of Figure \ref{f10}, which plots [O/Na] versus
[Al/H] (i.e., the elements exhibiting the largest abundance ranges).  In
general, each group spans a relatively small range in [O/Na] and [Al/H], and
the median [O/Na] ratio monotonically decreases as a function of [Al/H].
Additionally, the typical separation in median [O/Na] between adjacent 
populations is $\sim$0.35 dex, which is about two times larger than the 
dispersion within each group ($\sim$0.07 dex, on average).  Similarly, the
typical separation in median [Al/H] between adjacent groups is $\sim$0.3 dex,
which is about 25$\%$ larger than the typical [Al/H] dispersion within each 
group ($\sim$0.2 dex).

\subsection{Comparisons with Similar Clusters}

While nearly all monometallic clusters host 2--3 chemically distinct 
populations (e.g., Carretta et al. 2009a; Piotto et al. 2015; Milone et al. 
2017), those with $>$ 3 are relatively rare.  Among the monometallic clusters
studied in the literature, NGC 2808 is the only convincing case of a cluster
hosting at least five populations (Carretta 2015; Milone et al. 2015).  The 
UV--optical color magnitude diagrams of NGC 5986 are not as complex as those 
of NGC 2808 (e.g., see Piotto et al. 2015; their Figures 6 and 10), perhaps
because the abundance variations in NGC 5986 are not as extreme, but the 
data presented in Figures \ref{f8}--\ref{f10} support both clusters hosting a 
similar number of components.  An examination of the pseudo--color--magnitude 
diagram compilations in Piotto et al. (2015) and Milone et al. (2017) suggests 
that NGC 5986 is actually very similar to the cluster M 13.  Furthermore, the
pseudo--color--magnitude diagram of NGC 5986 also appears to contain more 
photometric sequences than the case of NGC 6752, which is confirmed to have at 
least three distinct populations (Carretta et al. 2012a), and is significantly 
more complex than a typical cluster like M 3.

\begin{figure}
\epsscale{1.00}
\plotone{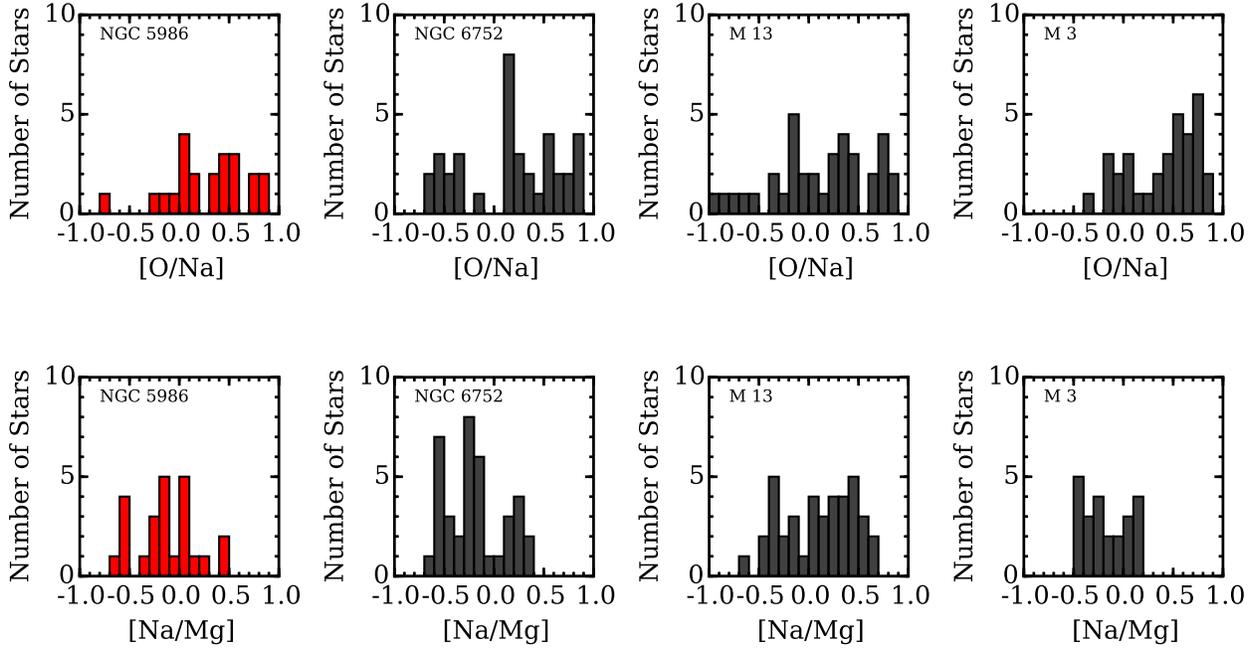}
\caption{\emph{Top:} Histograms of the [O/Na] distributions for NGC 5986, NGC
6752, M 13, and M 3 are shown with 0.1 dex bins.  Note that all four similar
metallicity clusters exhibit multimodal distributions.  \emph{Bottom:} Similar 
plots illustrating the [Na/Mg] distributions for the same clusters.  The data 
for NGC 6752, M 13, and M 3 are from Yong et al. (2005), Sneden et al. (2004), 
and Cohen \& Mel{\'e}ndez (2005), and have been shifted to have approximately 
the same maximum [O/Fe], maximum [Mg/Fe], minimum [Na/Fe], and minimum [Al/Fe] 
abundances as NGC 5986.  The red histograms indicate the data obtained for 
this paper.}
\label{f11}
\end{figure}

In Figure \ref{f11}, we compare histograms of the [O/Na] and [Na/Mg] 
distributions for NGC 5986 and the similar metallicity clusters M 13, NGC 6752,
and M 3.  Figure \ref{f11} reinforces the conclusions from Section 5.4 that
NGC 5986 hosts 4--5 populations, and also shows that the presence of multiple
distinct populations can be easily detected in the three other clusters as 
well.  Specifically, Figure \ref{f11} indicates that NGC 6752 has at least 
3 populations, M 3 has 2--3 populations, and M 13 has 4--5 populations (see
also Monelli et al. 2013).  

\begin{figure}
\epsscale{1.00}
\plotone{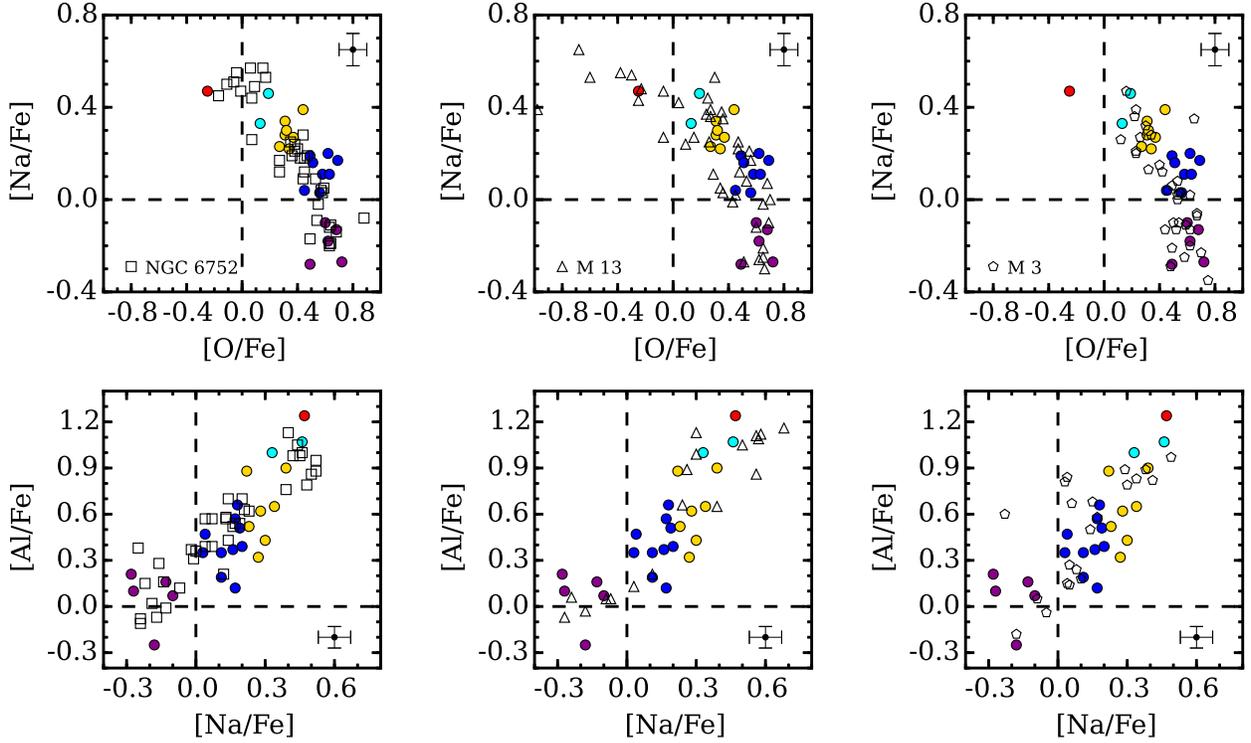}
\caption{The [O/Fe], [Na/Fe], and [Al/Fe] abundance patterns of NGC 5986 (same
colored symbols as in Figure \ref{f8}) are compared with those of the similar 
metallicity globular clusters NGC 6752 (left; open boxes), M 13 (middle; open
triangles), and M 3 (right; open pentagons).  The data for NGC 6752, M 13, and 
M 3 are from the same sources as in Figure \ref{f11}.  Similarly, the literature
data have been shifted to have approximately the same maximum [O/Fe], minimum 
[Na/Fe], and minimum [Al/Fe] abundances as NGC 5986.}
\label{f12}
\end{figure}

\begin{figure}
\epsscale{1.00}
\plotone{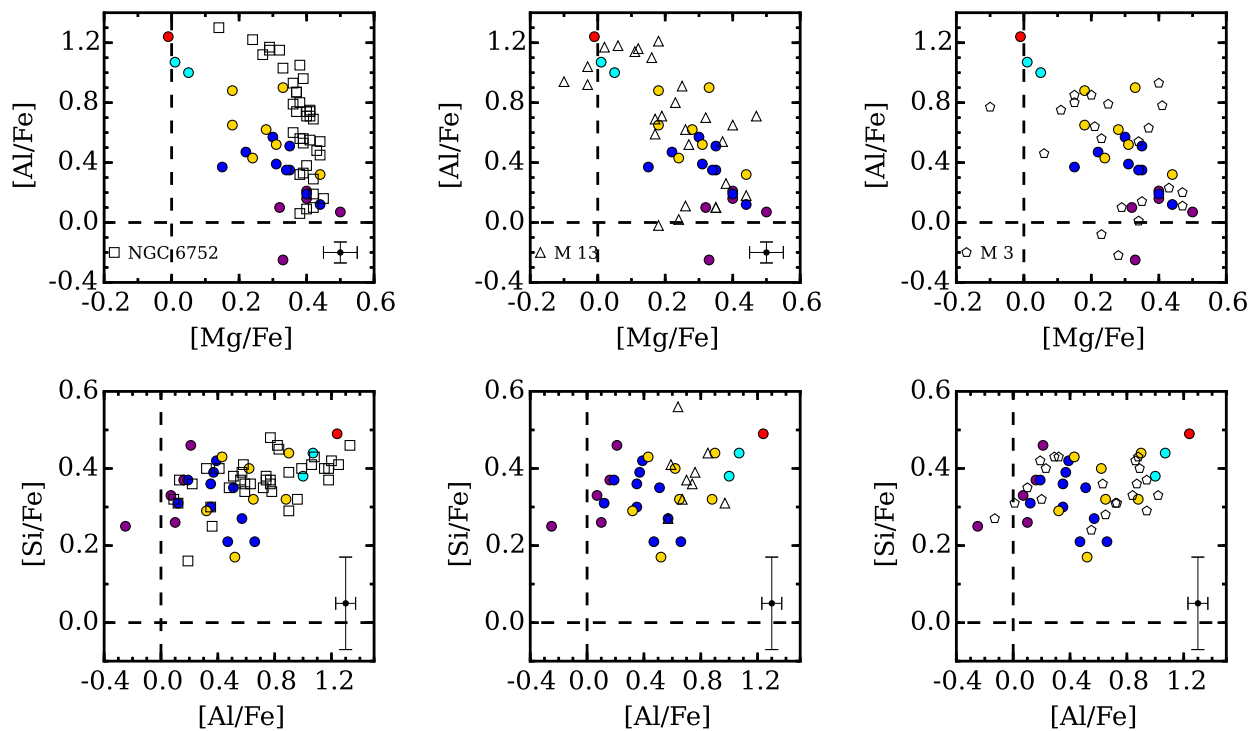}
\caption{The [Mg/Fe], [Al/Fe], and [Si/Fe] abundances of NGC 5986 are compared 
with those of the similar metallicity globular clusters NGC 6752 (left), M 13 
(middle), and M 3 (right).  Similar to Figure \ref{f12}, the abundances of 
NGC 6752, M 13, and M 3 have been shifted to have approximately the same
maximum [Mg/Fe], minimum [Al/Fe], and minimum [Si/Fe] abundances as NGC 5986.
The colors and symbols are the same as those in Figures \ref{f8} and \ref{f12}.}
\label{f13}
\end{figure}

The similar enrichment histories of NGC 5986 and M 13 are further solidified
in Figures \ref{f12}--\ref{f13} where we directly compare the O--Na, Na--Al,
Mg--Al, and Si--Al distributions of all four clusters\footnote{For Figures
\ref{f11}--\ref{f13}, we have applied systematic offsets to the literature 
[X/Fe] abundances so that the equivalent \emph{P1} populations of each cluster
match the \emph{P1} composition of NGC 5986.}.  Although NGC 5986 and NGC 6752 
share similar ranges in [O/Fe], [Na/Fe], [Al/Fe], and [Si/Fe], the full range 
in [Mg/Fe] is smaller for NGC 6752 and the two clusters exhibit different 
Mg--Al distributions.  Similarly, the \emph{P1}, \emph{P2}, and \emph{I1} 
populations of NGC 5986 almost identically match the M 3 distribution, but M 3 
does not possess stars matching the \emph{I2} and \emph{E} chemical 
compositions.  In contrast, NGC 5986 and M 13 exhibit nearly identical light 
element distributions, with the main difference being that the equivalent 
\emph{E} population in M 13 may extend to somewhat lower values of [O/Fe] and 
higher values of [Na/Fe].  We conclude that NGC 5986 and M 13 likely shared 
similar chemical enrichment histories and were probably enriched by similar
classes of polluters.

\subsection{Spatial Mixing of First and Second Generation Stars}

A commonly adopted globular cluster formation model posits that stars with
compositions similar to the \emph{P1} and \emph{P2} groups in NGC 5986 are the
first to form and that these stars are initially distributed at all cluster
radii (e.g., D'Ercole et al. 2008).  Subsequently, low velocity gas ejected 
from sources such as intermediate mass AGB stars forms a cooling flow that 
is funneled toward the cluster core where a second, more centrally 
concentrated, population forms.  As a result, stars with compositions similar
to the \emph{I1}, \emph{I2}, and \emph{E} populations of NGC 5986 initially 
have radial distributions that are distinct from those of first generation
stars.  In this model, a cluster's dynamical evolution favors the preferential 
loss of outer first generation stars, and remnants of the cluster's initial
population gradient may still be observable.  However, some simulations 
indicate that first and second generation stars may become spatially mixed 
after $>$ 60$\%$ of the initial cluster mass has been lost (e.g., Vesperini et 
al. 2013; Miholics et al. 2015).

An examination of the radial distributions of first and second generation stars
in monometallic Galactic globular clusters has produced conflicting results.  
For example, several earlier papers found that a large fraction of clusters 
tended to have centrally concentrated second generation populations (e.g.,
Carretta et al. 2009a, 2010c; Kravtsov et al. 2011; Lardo et al. 2011; Nataf 
et al. 2011; Johnson \& Pilachowski 2012; Milone et al. 2012a; Richer et al. 
2013; Cordero et al. 2014; Massari et al. 2016; Simioni et al. 2016).  However,
similar studies have also argued that many clusters either have no radial 
gradient (e.g., Iannicola et al. 2009; Lardo et al. 2011; Milone et al. 2013; 
Dalessando et al. 2014; Cordero et al. 2015; Nardiello et al. 2015; Vanderbeke 
et al. 2015) or that the first generation stars are actually the most centrally
concentrated (e.g., Larsen et al. 2015; Vanderbeke et al. 2015; Lim et al. 
2016).  Although the measurements are typically straight--forward, biases can 
be introduced because clusters are at different stages in their dynamical 
evolution and various studies frequently sample different cluster regions.  

\begin{figure}
\epsscale{1.00}
\plotone{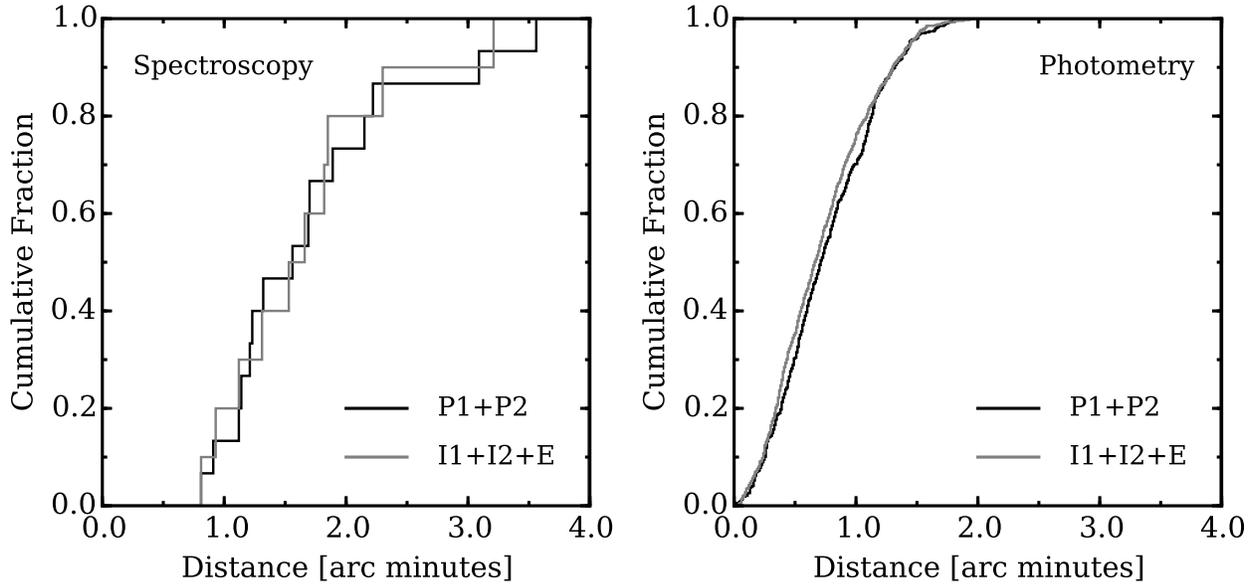}
\caption{\emph{Left:} The cumulative distribution functions of the \emph{P1} 
and \emph{P2} populations (black line) and \emph{I1}, \emph{I2}, and \emph{E} 
populations (grey line), identified via spectroscopy, are plotted as a function
of the projected radial distance from the cluster center.  \emph{Right:} A 
similar plot indicating the radial distributions of the same populations 
inferred from the pseudo--colors shown in Figure \ref{f9}.  For the photometric
data, we divided the RGB sample in half and assumed that the stars with bluer
pseudo--colors are predominantly associated with the \emph{I1}, \emph{I2}, 
and \emph{E} populations while those with redder pseudo--colors are
predominantly associated with the \emph{P1} and \emph{P2} populations.  Note
that we have adopted a half--mass radius of 0.98$\arcmin$ (Harris 1996) for
NGC 5986.}
\label{f14}
\end{figure}

In Figure \ref{f14}, we show the cumulative radial distributions of the first 
(\emph{P1} $+$ \emph{P2}) and second (\emph{I1} $+$ \emph{I2} $+$ \emph{E}) 
generation stars measured via spectroscopy and inferred from the 
\emph{HST} photometry in Figure \ref{f10}.  The advantages of examining
both data sets include a substantial increase in the sample size and an 
extension of the radial coverage from $\sim$0.8--3.6 half--mass 
radii\footnote{We have adopted 0.98$\arcmin$, the projected half--light radius 
listed in Harris (1996), as the cluster's half--mass radius.  We are assuming 
that the half--light and half--mass radii are approximately equal.} to 
$\sim$0.01--3.6 half--mass radii.  The additional coverage is especially 
important because models from Vesperini et al. (2013) indicate that the region 
between $\sim$1--2 half--mass radii may be where the local ratio of 
second--to--first generation stars is equivalent to the global ratio.  

Interestingly, both the spectroscopic and photometric data in Figure \ref{f14} 
show that the first and second generation stars in NGC 5986 share similar
radial distributions from the core out to more than 3.5 half--mass radii.  
Two--sided Kolmogorov--Smirnov tests of the spectroscopic and photometric sets 
provide \emph{p}--values of 0.999 and 0.175, respectively, which indicate that
the data do not provide enough evidence to reject the null hypothesis that the 
first and second generation stars were drawn from the same parent distribution.
Therefore, we conclude that the various stellar populations in NGC 5986 are 
largely spatially mixed over a wide range in cluster radii.  
If we assume that the first and second generation stars were originally 
segregated but are now spatially mixed, it is possible that the cluster may 
have lost at least 60--80$\%$ of its original mass (Vesperini et 
al. 2013; Miholics et al. 2015), perhaps as a consequence of its highly 
eccentric inner Galaxy orbit (Casetti--Dinescu et al. 2007).  However, we 
caution that the radial distributions may be more nuanced, and it may not 
necessarily be appropriate to analyze multiple groups (e.g., \emph{I1}, 
\emph{I2}, and \emph{E}) as a single unit.  For example, at least in M 13
Johnson \& Pilachowski (2012) showed that only the most O--poor stars (\emph{E}
group) may be centrally concentrated, and Cordero et al. (2017) further 
noted that the \emph{E} population exhibits faster rotation than the other
sub--populations.

\section{SUMMARY}

This paper utilizes high resolution, high S/N data from the 
\emph{Magellan}--M2FS instrument to obtain radial velocities and chemical 
abundances for a sample of 43 potential RGB and AGB stars near the Galactic
globular cluster NGC 5986.  A combination of velocity and [Fe/H] measurements 
identified 27/43 stars in our sample as likely cluster members, but we were 
only able to measure detailed abundances for 25/27 member stars.  We found
NGC 5986 to have a mean heliocentric radial velocity of $+$99.76 km 
s$^{\rm -1}$ ($\sigma$ = 7.44 km s$^{\rm -1}$) and a mean metallicity of 
[Fe/H] = --1.54 dex ($\sigma$ = 0.08 dex).

The cluster's overall chemical composition characteristics are comparable to
other similar metallicity clusters.  For example, the heavier 
$\alpha$--elements are uniformly enhanced with $\langle$[Si,Ca/Fe]$\rangle$ = 
$+$0.30 dex ($\sigma$ = 0.08 dex) and the Fe--peak elements all exhibit 
nearly solar [X/Fe] ratios.  Interestingly, the neutron--capture elements are
moderately enhanced with $\langle$[La/Fe]$\rangle$ = $+$0.42 dex ($\sigma$ = 
0.11 dex) and $\langle$[Eu/Fe]$\rangle$ = $+$0.76 dex ($\sigma$ = 0.08 dex),
and NGC 5986 may be among the most Eu--rich clusters known in the Galaxy.  
The Eu enhancements are comparable to those found in similar metallicity stars 
in several Local Group dwarf galaxies, but NGC 5986 does not share the trait
of exhibiting a large [Eu/Fe] dispersion.  The cluster's low [La/Eu] ratios
combined with its small [Fe/H] and [La/Eu] dispersions preclude NGC 5986 from 
being a member of the iron--complex class, which is characterized by having 
large dispersions in [La/Eu] that are correlated with metallicity spreads.  

We find that NGC 5986 exhibits all of the classical light element abundance 
relations, including strong anti--correlations between O--Na, Na--Mg, and 
Mg--Al and correlations between Na--Al and Al--Si.  The combined presence of 
a Mg--Al anti--correlation and Al--Si correlation suggests that the gas from
which the second generation stars formed must have experienced burning 
temperatures of at least 65--70 MK.  However, the cluster does not exhibit
very low [Mg/Fe] ratios nor a particularly large range in [Ca/Mg] that would 
indicate significant processing at temperatures exceeding $\sim$100 MK.
The abundance anti--correlations are also not well--fit by a simple dilution 
model, which suggests that more than one class of polluters likely contributed 
to the cluster's self--enrichment.  

One of the most striking results is the discrete nature of the light element
abundance patterns, and we find evidence that NGC 5986 may host at least 4--5
different populations with distinct compositions.  Although the sample sizes
within each sub--population are small, statistical tests support the idea that
the commonly observed ``primordial" and ``intermediate" groups in NGC 5986 may
each be further decomposed into populations with distinctly different light
element compositions.  Our analysis also identified at least one star
that is very O/Mg--poor and Na/Al/Si--rich, which we designated as an 
``extreme" (\emph{E}) population member.  We find that the two primordial 
(\emph{P1} and \emph{P2}) and intermediate (\emph{I1} and \emph{I2}) groups
are present in the proportions 20$\%$, 40$\%$, 28$\%$, and 8$\%$, respectively,
while the single \emph{E} population star constitutes the remaining 4$\%$ of 
our sample.  If confirmed via photometry and/or larger sample spectroscopic
analyses, NGC 5986 would join NGC 2808 as the only known monometallic clusters 
that host $>$ 3 distinct populations.  However, we note that a comparison
between NGC 5986 and M 13 revealed that the two clusters exhibit almost 
identical composition patterns, and in fact M 13 may also be composed of at 
least four different populations.

Interestingly, an examination of the radial distributions of first (\emph{P1} 
and \emph{P2}) and second (\emph{I1}, \emph{I2}, and \emph{E}) generation stars
in NGC 5986 suggests that the populations are well--mixed within the cluster.
This result is seemingly confirmed from both the spectroscopic sample analyzed
here and a photometric sample obtained from the literature.  If confirmed, 
the full spatial mixing of NGC 5986's various populations may suggest either
that the stars were never radially segregated or that the cluster may have 
lost $\sim$60--80$\%$ of its original mass.

\acknowledgements

This research has made use of NASA's Astrophysics Data System Bibliographic
Services.  This publication has made use of data products from the Two Micron
All Sky Survey, which is a joint project of the University of Massachusetts
and the Infrared Processing and Analysis Center/California Institute of
Technology, funded by the National Aeronautics and Space Administration and
the National Science Foundation.  C.I.J. gratefully acknowledges support from
the Clay Fellowship, administered by the Smithsonian Astrophysical Observatory.
M.M. is grateful for support from the National Science Foundation to develop 
M2FS (AST--0923160) and carry out the observations reported here 
(AST--1312997), and to the University of Michigan for its direct support of 
M2FS construction and operation.  M.G.W. is supported by National Science 
Foundation grants AST--1313045 and AST--1412999.  R.M.R acknowledges support 
from grant AST--1413755 from the National Science Foundation.  E.W.O. 
acknowledges support from the National Science Foundation under grant 
AST--1313006.

\clearpage

\clearpage

\clearpage

\clearpage

\clearpage

\clearpage

\clearpage

\clearpage

\clearpage

\clearpage

\clearpage

\clearpage

\clearpage

\clearpage

\clearpage
\setlength{\hoffset}{-0.50in}
\tablenum{1}
\tablecolumns{12}
\tablewidth{0pt}

\begin{deluxetable}{cccccccccccc}
\tabletypesize{\tiny}
\tablecaption{Star Identifiers, Coordinates, Photometry, and Velocities}
\tablehead{
\colhead{Star Name}	&
\colhead{Alt. ID\tablenotemark{a}}      &
\colhead{RA}      &
\colhead{DEC}      &
\colhead{B}      &
\colhead{V}      &
\colhead{J}      &
\colhead{H}      &
\colhead{K$_{\rm S}$}      &
\colhead{RV$_{\rm helio.}$}      &
\colhead{RV$_{\rm helio.}$ Error}	&
\colhead{Evol. State}	\\
\colhead{(2MASS)}      &
\colhead{}      &
\colhead{(degrees)}      &
\colhead{(degrees)}      &
\colhead{(mag.)}      &
\colhead{(mag.)}      &
\colhead{(mag.)}      &
\colhead{(mag.)}      &
\colhead{(mag.)}      &
\colhead{(km s$^{\rm -1}$)}      &
\colhead{(km s$^{\rm -1}$)}      &
\colhead{}
}

\startdata
\hline
\multicolumn{12}{c}{Cluster Members}       \\
\hline
15455510$-$3746063	&	153	&	236.479603	&	$-$37.768421	&	16.396	&	15.095	&	12.406	&	11.750	&	11.597	&	104.99	&	0.42	&	RGB	\\
15455524$-$3746528	&	55	&	236.480173	&	$-$37.781353	&	15.646	&	14.155	&	11.277	&	10.568	&	10.371	&	107.77	&	0.21	&	AGB	\\
15455531$-$3748266	&	85	&	236.480472	&	$-$37.807415	&	15.816	&	14.481	&	11.812	&	11.145	&	11.008	&	95.30	&	0.33	&	AGB	\\
15455682$-$3747414	&	83	&	236.486777	&	$-$37.794834	&	15.967	&	14.471	&	11.555	&	10.837	&	10.655	&	100.24	&	0.24	&	RGB	\\
15455728$-$3748245	&	90	&	236.488681	&	$-$37.806828	&	16.018	&	14.560	&	11.686	&	10.940	&	10.776	&	111.04	&	0.28	&	RGB	\\
15455743$-$3745416	&	80	&	236.489299	&	$-$37.761562	&	15.871	&	14.450	&	11.635	&	10.909	&	10.698	&	102.67	&	0.21	&	RGB	\\
15455768$-$3746466	&	111	&	236.490334	&	$-$37.779633	&	16.147	&	14.786	&	12.035	&	11.347	&	11.196	&	93.90	&	0.23	&	RGB	\\
15455843$-$3747538	&	81	&	236.493468	&	$-$37.798294	&	15.927	&	14.452	&	11.575	&	10.844	&	10.655	&	92.88	&	0.19	&	RGB	\\
15455886$-$3747091	&	127	&	236.495254	&	$-$37.785885	&	16.161	&	14.944	&	12.367	&	11.742	&	11.534	&	102.25	&	0.35	&	AGB	\\
15455926$-$3746383	&	77	&	236.496932	&	$-$37.777321	&	15.884	&	14.414	&	11.475	&	10.780	&	10.584	&	90.17	&	0.24	&	RGB	\\
15460024$-$3748232	&	25	&	236.501022	&	$-$37.806450	&	15.167	&	13.679	&	10.888	&	10.214	&	10.022	&	101.60	&	0.22	&	Post$-$AGB?	\\
15460030$-$3746057	&	78	&	236.501274	&	$-$37.768265	&	15.854	&	14.419	&	11.562	&	10.884	&	10.679	&	83.96	&	0.24	&	RGB	\\
15460078$-$3745426	&	86	&	236.503273	&	$-$37.761837	&	15.663	&	14.496	&	11.714	&	11.055	&	10.869	&	103.40	&	0.24	&	Post$-$AGB?	\\
15460253$-$3746035	&	132	&	236.510547	&	$-$37.767647	&	16.151	&	14.968	&	12.477	&	11.886	&	11.737	&	84.93	&	0.33	&	AGB	\\
15460303$-$3745286	&	45	&	236.512634	&	$-$37.757965	&	15.553	&	14.029	&	11.112	&	10.418	&	10.231	&	110.23	&	0.19	&	AGB	\\
15460317$-$3747598	&	126	&	236.513216	&	$-$37.799950	&	16.269	&	14.943	&	12.326	&	11.676	&	11.501	&	93.68	&	0.33	&	RGB	\\
15460332$-$3748249	&	99	&	236.513847	&	$-$37.806942	&	16.073	&	14.676	&	11.921	&	11.184	&	11.040	&	96.90	&	0.21	&	RGB	\\
15460476$-$3749186	&	112	&	236.519865	&	$-$37.821861	&	16.144	&	14.787	&	12.038	&	11.365	&	11.188	&	101.56	&	0.24	&	RGB	\\
15460690$-$3750184	&	152	&	236.528773	&	$-$37.838455	&	16.434	&	15.082	&	12.480	&	11.700	&	11.538	&	98.94	&	0.36	&	RGB	\\
15460751$-$3746542	&	60	&	236.531298	&	$-$37.781746	&	15.674	&	14.221	&	11.387	&	10.683	&	10.458	&	113.51	&	0.24	&	AGB	\\
15460895$-$3749046	&	57	&	236.537307	&	$-$37.817947	&	15.748	&	14.178	&	11.174	&	10.418	&	10.247	&	101.88	&	0.26	&	RGB	\\
15460957$-$3747333	&	102	&	236.539912	&	$-$37.792603	&	15.951	&	14.698	&	12.108	&	11.477	&	11.366	&	97.21	&	0.43	&	AGB	\\
15461022$-$3749558	&	116	&	236.542585	&	$-$37.832172	&	16.222	&	14.852	&	12.066	&	11.381	&	11.221	&	94.54	&	0.28	&	RGB	\\
15461025$-$3746168	&	123	&	236.542741	&	$-$37.771343	&	16.225	&	14.924	&	12.318	&	11.635	&	11.506	&	107.23	&	0.25	&	RGB	\\
15461222$-$3747118	&	140	&	236.550938	&	$-$37.786636	&	16.270	&	15.001	&	12.396	&	11.717	&	11.562	&	108.25	&	0.34	&	RGB	\\
15461303$-$3746009	&	129	&	236.554316	&	$-$37.766926	&	16.278	&	14.958	&	12.264	&	11.627	&	11.448	&	97.89	&	0.25	&	RGB	\\
15461630$-$3744468	&	65	&	236.567931	&	$-$37.746349	&	15.849	&	14.290	&	11.238	&	10.460	&	10.337	&	96.71	&	0.21	&	RGB	\\
\hline
\multicolumn{12}{c}{Non--Members}       \\
\hline
15452886$-$3748562	&	270	&	236.370278	&	$-$37.815613	&	17.352	&	15.660	&	12.433	&	11.811	&	11.598	&	28.98	&	1.25	&	\nodata	\\
15453279$-$3743579	&	176	&	236.386664	&	$-$37.732777	&	17.054	&	15.214	&	11.784	&	10.945	&	10.717	&	$-$1.95	&	0.39	&	\nodata	\\
15453914$-$3752110	&	135	&	236.413115	&	$-$37.869740	&	16.561	&	14.972	&	12.285	&	11.683	&	11.548	&	8.96	&	0.58	&	\nodata	\\
15454023$-$3741589	&	87	&	236.417660	&	$-$37.699699	&	15.916	&	14.497	&	11.822	&	11.114	&	10.978	&	$-$10.28	&	0.39	&	\nodata	\\
15455164$-$3747031	&	158	&	236.465173	&	$-$37.784206	&	16.779	&	15.109	&	12.301	&	11.696	&	11.491	&	105.81	&	1.06	&	\nodata	\\
15460039$-$3747517	&	110	&	236.501659	&	$-$37.797695	&	15.988	&	14.784	&	12.330	&	11.742	&	11.568	&	$-$46.47	&	0.24	&	\nodata	\\
15460184$-$3749195	&	151	&	236.507702	&	$-$37.822098	&	16.375	&	15.081	&	12.397	&	11.756	&	11.558	&	$-$38.94	&	0.44	&	\nodata	\\
15460206$-$3745372	&	69	&	236.508607	&	$-$37.760357	&	15.728	&	14.319	&	11.579	&	10.853	&	10.713	&	146.82	&	0.43	&	\nodata	\\
15460582$-$3748002	&	121	&	236.524272	&	$-$37.800072	&	16.098	&	14.907	&	12.389	&	11.792	&	11.647	&	46.02	&	0.54	&	\nodata	\\
15460612$-$3749563	&	100	&	236.525510	&	$-$37.832314	&	16.159	&	14.684	&	11.714	&	11.058	&	10.861	&	$-$36.50	&	0.54	&	\nodata	\\
15461993$-$3743446	&	139	&	236.583058	&	$-$37.729076	&	16.248	&	14.994	&	12.452	&	11.824	&	11.721	&	$-$28.73	&	0.31	&	\nodata	\\
15462147$-$3744159	&	194	&	236.589484	&	$-$37.737766	&	16.735	&	15.282	&	12.403	&	11.641	&	11.509	&	8.47	&	0.29	&	\nodata	\\
15462221$-$3741066	&	71	&	236.592569	&	$-$37.685184	&	16.030	&	14.363	&	11.231	&	10.471	&	10.277	&	$-$2.99	&	0.49	&	\nodata	\\
15462726$-$3744267	&	228	&	236.613608	&	$-$37.740757	&	17.119	&	15.482	&	12.293	&	11.462	&	11.231	&	$-$36.84	&	0.47	&	\nodata	\\
15463408$-$3749556	&	30	&	236.642030	&	$-$37.832115	&	15.200	&	13.724	&	11.012	&	10.330	&	10.169	&	$-$77.36	&	0.40	&	\nodata	\\
15464077$-$3744246	&	\nodata	&	236.669894	&	$-$37.740177	&	\nodata	&	\nodata	&	12.120	&	11.561	&	11.376	&	$-$72.71	&	0.38	&	\nodata	\\
\enddata

\tablenotetext{a}{Identifiers are from Alves et al. (2001).}

\end{deluxetable}

\clearpage
\tablenum{2}
\tablecolumns{5}
\tablewidth{0pt}

\begin{deluxetable}{ccccc}
\tablecaption{Model Atmosphere Parameters}
\tablehead{
\colhead{Star Name}     &
\colhead{T$_{\rm eff}$} &
\colhead{log(g)}      &
\colhead{[Fe/H]}      &
\colhead{$\xi$$_{\rm mic.}$}      \\
\colhead{(2MASS)}       &
\colhead{(K)}      &
\colhead{(cgs)}      &
\colhead{(dex)}      &
\colhead{(km s$^{\rm -1}$)}
}

\startdata
15455510$-$3746063	&	4475	&	1.15	&	$-$1.59	&	1.70	\\
15455524$-$3746528	&	4275	&	0.60	&	$-$1.60	&	1.90	\\
15455531$-$3748266	&	\nodata	&	\nodata	&	\nodata	&	\nodata	\\
15455682$-$3747414	&	4300	&	0.80	&	$-$1.57	&	1.80	\\
15455728$-$3748245	&	4300	&	0.80	&	$-$1.49	&	1.65	\\
15455743$-$3745416	&	4375	&	1.10	&	$-$1.46	&	1.80	\\
15455768$-$3746466	&	4375	&	0.90	&	$-$1.65	&	1.80	\\
15455843$-$3747538	&	4375	&	1.25	&	$-$1.52	&	1.95	\\
15455886$-$3747091	&	4600	&	1.25	&	$-$1.58	&	1.65	\\
15455926$-$3746383	&	4350	&	1.15	&	$-$1.42	&	1.75	\\
15460024$-$3748232	&	4300	&	0.45	&	$-$1.66	&	1.80	\\
15460030$-$3746057	&	4325	&	1.00	&	$-$1.52	&	1.80	\\
15460078$-$3745426	&	4325	&	0.90	&	$-$1.57	&	1.50	\\
15460253$-$3746035	&	4500	&	0.95	&	$-$1.71	&	1.55	\\
15460303$-$3745286	&	4250	&	0.50	&	$-$1.62	&	1.95	\\
15460317$-$3747598	&	4600	&	1.45	&	$-$1.46	&	1.60	\\
15460332$-$3748249	&	4350	&	1.05	&	$-$1.63	&	1.85	\\
15460476$-$3749186	&	4450	&	1.25	&	$-$1.50	&	1.65	\\
15460690$-$3750184	&	4400	&	0.95	&	$-$1.61	&	1.65	\\
15460751$-$3746542	&	4325	&	1.05	&	$-$1.52	&	1.85	\\
15460895$-$3749046	&	4225	&	0.65	&	$-$1.53	&	1.95	\\
15460957$-$3747333	&	\nodata	&	\nodata	&	\nodata	&	\nodata	\\
15461022$-$3749558	&	4550	&	1.50	&	$-$1.45	&	1.85	\\
15461025$-$3746168	&	4550	&	1.50	&	$-$1.45	&	1.60	\\
15461222$-$3747118	&	4550	&	1.60	&	$-$1.49	&	1.70	\\
15461303$-$3746009	&	4550	&	1.50	&	$-$1.49	&	1.80	\\
15461630$-$3744468	&	4250	&	0.80	&	$-$1.53	&	2.00	\\
\enddata

\end{deluxetable}

\clearpage
\tablenum{3}
\tablecolumns{13}
\tablewidth{0pt}

\begin{deluxetable}{ccccccccccccc}
\rotate
\tabletypesize{\tiny}
\tablecaption{Chemical Abundances and Uncertainties: O--Ca}
\tablehead{
\colhead{Star Name}     &
\colhead{[O I/Fe]}	&
\colhead{$\Delta$[O I/Fe]}        &
\colhead{[Na I/Fe]}        &
\colhead{$\Delta$[Na I/Fe]}        &
\colhead{[Mg I/Fe]}        &
\colhead{$\Delta$[Mg I/Fe]}        &
\colhead{[Al I/Fe]}        &
\colhead{$\Delta$[Al I/Fe]}        &
\colhead{[Si I/Fe]}        &
\colhead{$\Delta$[Si I/Fe]}        &
\colhead{[Ca I/Fe]}        &
\colhead{$\Delta$[Ca I/Fe]}        \\
\colhead{(2MASS)}     &
\colhead{(dex)}	&
\colhead{(dex)} &
\colhead{(dex)} &
\colhead{(dex)} &
\colhead{(dex)} &
\colhead{(dex)} &
\colhead{(dex)} &
\colhead{(dex)} &
\colhead{(dex)} &
\colhead{(dex)} &
\colhead{(dex)} &
\colhead{(dex)} 
}

\startdata
15455510$-$3746063	&	0.62	&	0.10	&	0.20	&	0.09	&	0.31	&	0.05	&	0.39	&	0.07	&	0.42	&	0.13	&	0.31	&	0.07	\\
15455524$-$3746528	&	0.45	&	0.10	&	0.04	&	0.12	&	0.22	&	0.04	&	0.47	&	0.11	&	0.21	&	0.11	&	0.26	&	0.07	\\
15455531$-$3748266	&	\nodata	&	\nodata	&	\nodata	&	\nodata	&	\nodata	&	\nodata	&	\nodata	&	\nodata	&	\nodata	&	\nodata	&	\nodata	&	\nodata	\\
15455682$-$3747414	&	$-$0.25	&	0.10	&	0.47	&	0.06	&	$-$0.01	&	0.07	&	1.24	&	0.06	&	0.49	&	0.11	&	0.32	&	0.05	\\
15455728$-$3748245	&	0.27	&	0.10	&	0.23	&	0.04	&	0.31	&	0.07	&	0.52	&	0.13	&	0.17	&	0.15	&	0.24	&	0.06	\\
15455743$-$3745416	&	0.34	&	0.10	&	0.22	&	0.14	&	0.18	&	0.06	&	0.88	&	0.07	&	0.32	&	0.11	&	0.31	&	0.06	\\
15455768$-$3746466	&	0.60	&	0.10	&	$-$0.10	&	0.08	&	0.50	&	0.03	&	0.07	&	0.06	&	0.33	&	0.12	&	0.33	&	0.06	\\
15455843$-$3747538	&	0.69	&	0.10	&	0.17	&	0.03	&	0.44	&	0.05	&	0.12	&	0.07	&	0.31	&	0.11	&	0.28	&	0.05	\\
15455886$-$3747091	&	0.51	&	0.10	&	0.16	&	0.06	&	0.15	&	0.05	&	0.37	&	0.07	&	0.39	&	0.12	&	0.24	&	0.06	\\
15455926$-$3746383	&	0.63	&	0.10	&	0.11	&	0.08	&	0.34	&	0.05	&	0.35	&	0.07	&	0.30	&	0.12	&	0.30	&	0.05	\\
15460024$-$3748232	&	0.37	&	0.10	&	0.27	&	0.04	&	0.44	&	0.05	&	0.32	&	0.07	&	0.29	&	0.11	&	0.28	&	0.07	\\
15460030$-$3746057	&	0.49	&	0.10	&	0.19	&	0.03	&	0.35	&	0.05	&	0.51	&	0.07	&	0.35	&	0.18	&	0.25	&	0.05	\\
15460078$-$3745426	&	0.68	&	0.10	&	$-$0.13	&	0.19	&	0.40	&	0.03	&	0.16	&	0.07	&	0.37	&	0.11	&	0.26	&	0.06	\\
15460253$-$3746035	&	0.32	&	0.10	&	0.30	&	0.08	&	0.24	&	0.05	&	0.43	&	0.10	&	0.43	&	0.12	&	0.15	&	0.06	\\
15460303$-$3745286	&	0.62	&	0.10	&	$-$0.18	&	0.06	&	0.33	&	0.04	&	$-$0.25	&	0.05	&	0.25	&	0.11	&	0.29	&	0.06	\\
15460317$-$3747598	&	0.44	&	0.10	&	0.39	&	0.11	&	0.33	&	0.05	&	0.90	&	0.05	&	0.44	&	0.11	&	0.22	&	0.06	\\
15460332$-$3748249	&	0.58	&	0.10	&	0.11	&	0.07	&	0.40	&	0.03	&	0.19	&	0.07	&	0.37	&	0.11	&	0.30	&	0.06	\\
15460476$-$3749186	&	0.72	&	0.10	&	$-$0.27	&	0.09	&	0.32	&	0.05	&	0.10	&	0.07	&	0.26	&	0.12	&	0.19	&	0.06	\\
15460690$-$3750184	&	0.31	&	0.10	&	0.28	&	0.03	&	0.28	&	0.05	&	0.62	&	0.05	&	0.40	&	0.11	&	0.29	&	0.06	\\
15460751$-$3746542	&	0.31	&	0.10	&	0.34	&	0.08	&	0.18	&	0.05	&	0.65	&	0.07	&	0.32	&	0.12	&	0.27	&	0.06	\\
15460895$-$3749046	&	0.49	&	0.10	&	$-$0.28	&	0.08	&	0.40	&	0.05	&	0.21	&	0.07	&	0.46	&	0.12	&	0.33	&	0.06	\\
15460957$-$3747333	&	\nodata	&	\nodata	&	\nodata	&	\nodata	&	\nodata	&	\nodata	&	\nodata	&	\nodata	&	\nodata	&	\nodata	&	\nodata	&	\nodata	\\
15461022$-$3749558	&	\nodata	&	\nodata	&	0.17	&	0.03	&	0.30	&	0.05	&	0.57	&	0.07	&	0.27	&	0.11	&	0.21	&	0.05	\\
15461025$-$3746168	&	\nodata	&	\nodata	&	0.18	&	0.03	&	\nodata	&	\nodata	&	0.66	&	0.04	&	0.21	&	0.11	&	0.22	&	0.06	\\
15461222$-$3747118	&	0.13	&	0.10	&	0.33	&	0.03	&	0.05	&	0.05	&	1.00	&	0.10	&	0.38	&	0.12	&	0.26	&	0.06	\\
15461303$-$3746009	&	0.19	&	0.10	&	0.46	&	0.08	&	0.01	&	0.05	&	1.07	&	0.07	&	0.44	&	0.12	&	0.26	&	0.06	\\
15461630$-$3744468	&	0.56	&	0.10	&	0.03	&	0.08	&	0.35	&	0.05	&	0.35	&	0.05	&	0.36	&	0.11	&	0.30	&	0.06	\\
\enddata

\end{deluxetable}

\clearpage
\tablenum{4}
\tablecolumns{13}
\tablewidth{0pt}

\begin{deluxetable}{ccccccccccccc}
\rotate
\tabletypesize{\tiny}
\tablecaption{Chemical Abundances and Uncertainties: Cr--Eu}
\tablehead{
\colhead{Star Name}     &
\colhead{[Cr I/Fe]}	&
\colhead{$\Delta$[Cr I/Fe]}        &
\colhead{[Fe I/H]}        &
\colhead{$\Delta$[Fe I/H]}        &
\colhead{[Fe II/H]}        &
\colhead{$\Delta$[Fe II/H]}        &
\colhead{[Ni I/Fe]}        &
\colhead{$\Delta$[Ni I/Fe]}        &
\colhead{[La II/Fe]}        &
\colhead{$\Delta$[La II/Fe]}        &
\colhead{[Eu II/Fe]}        &
\colhead{$\Delta$[Eu II/Fe]}        \\
\colhead{(2MASS)}     &
\colhead{(dex)}	&
\colhead{(dex)} &
\colhead{(dex)} &
\colhead{(dex)} &
\colhead{(dex)} &
\colhead{(dex)} &
\colhead{(dex)} &
\colhead{(dex)} &
\colhead{(dex)} &
\colhead{(dex)} &
\colhead{(dex)} &
\colhead{(dex)} 
}

\startdata
15455510$-$3746063	&	0.05	&	0.09	&	$-$1.59	&	0.10	&	$-$1.58	&	0.12	&	$-$0.14	&	0.07	&	0.52	&	0.10	&	0.68	&	0.10	\\
15455524$-$3746528	&	$-$0.08	&	0.07	&	$-$1.60	&	0.10	&	$-$1.60	&	0.12	&	$-$0.12	&	0.06	&	0.34	&	0.10	&	0.62	&	0.10	\\
15455531$-$3748266	&	\nodata	&	\nodata	&	\nodata	&	\nodata	&	\nodata	&	\nodata	&	\nodata	&	\nodata	&	\nodata	&	\nodata	&	\nodata	&	\nodata	\\
15455682$-$3747414	&	0.07	&	0.05	&	$-$1.57	&	0.10	&	$-$1.57	&	0.13	&	$-$0.16	&	0.06	&	0.38	&	0.10	&	0.67	&	0.10	\\
15455728$-$3748245	&	0.08	&	0.06	&	$-$1.49	&	0.10	&	$-$1.48	&	0.11	&	$-$0.14	&	0.06	&	0.24	&	0.10	&	0.76	&	0.11	\\
15455743$-$3745416	&	0.06	&	0.06	&	$-$1.46	&	0.10	&	$-$1.46	&	0.13	&	$-$0.14	&	0.06	&	0.51	&	0.10	&	0.89	&	0.11	\\
15455768$-$3746466	&	0.10	&	0.06	&	$-$1.65	&	0.10	&	$-$1.65	&	0.12	&	$-$0.02	&	0.06	&	0.23	&	0.10	&	0.70	&	0.15	\\
15455843$-$3747538	&	$-$0.02	&	0.05	&	$-$1.52	&	0.10	&	$-$1.51	&	0.12	&	$-$0.06	&	0.05	&	0.53	&	0.10	&	0.95	&	0.12	\\
15455886$-$3747091	&	\nodata	&	\nodata	&	$-$1.58	&	0.10	&	$-$1.58	&	0.11	&	$-$0.17	&	0.07	&	0.23	&	0.11	&	0.79	&	0.10	\\
15455926$-$3746383	&	0.02	&	0.05	&	$-$1.42	&	0.10	&	$-$1.42	&	0.13	&	$-$0.07	&	0.07	&	0.46	&	0.10	&	0.79	&	0.10	\\
15460024$-$3748232	&	\nodata	&	\nodata	&	$-$1.67	&	0.10	&	$-$1.64	&	0.12	&	$-$0.20	&	0.05	&	0.36	&	0.11	&	0.69	&	0.10	\\
15460030$-$3746057	&	0.07	&	0.05	&	$-$1.51	&	0.10	&	$-$1.52	&	0.11	&	$-$0.18	&	0.05	&	0.53	&	0.10	&	0.79	&	0.10	\\
15460078$-$3745426	&	$-$0.05	&	0.06	&	$-$1.58	&	0.10	&	$-$1.56	&	0.13	&	$-$0.14	&	0.05	&	0.35	&	0.15	&	0.69	&	0.10	\\
15460253$-$3746035	&	0.14	&	0.06	&	$-$1.72	&	0.10	&	$-$1.69	&	0.12	&	$-$0.13	&	0.06	&	0.53	&	0.11	&	0.68	&	0.10	\\
15460303$-$3745286	&	$-$0.03	&	0.06	&	$-$1.61	&	0.10	&	$-$1.62	&	0.12	&	$-$0.16	&	0.06	&	0.32	&	0.11	&	0.66	&	0.10	\\
15460317$-$3747598	&	0.02	&	0.06	&	$-$1.46	&	0.10	&	$-$1.46	&	0.13	&	0.03	&	0.15	&	0.34	&	0.15	&	0.82	&	0.10	\\
15460332$-$3748249	&	0.08	&	0.06	&	$-$1.63	&	0.10	&	$-$1.62	&	0.11	&	$-$0.17	&	0.07	&	0.49	&	0.10	&	0.74	&	0.10	\\
15460476$-$3749186	&	0.06	&	0.06	&	$-$1.50	&	0.10	&	$-$1.50	&	0.12	&	$-$0.12	&	0.06	&	0.48	&	0.14	&	0.76	&	0.11	\\
15460690$-$3750184	&	0.04	&	0.06	&	$-$1.61	&	0.10	&	$-$1.60	&	0.11	&	$-$0.15	&	0.10	&	0.29	&	0.11	&	0.71	&	0.10	\\
15460751$-$3746542	&	0.02	&	0.06	&	$-$1.51	&	0.10	&	$-$1.52	&	0.12	&	$-$0.14	&	0.05	&	0.43	&	0.10	&	0.81	&	0.10	\\
15460895$-$3749046	&	0.00	&	0.06	&	$-$1.53	&	0.10	&	$-$1.53	&	0.13	&	$-$0.10	&	0.07	&	0.35	&	0.10	&	0.75	&	0.10	\\
15460957$-$3747333	&	\nodata	&	\nodata	&	\nodata	&	\nodata	&	\nodata	&	\nodata	&	\nodata	&	\nodata	&	\nodata	&	\nodata	&	\nodata	&	\nodata	\\
15461022$-$3749558	&	0.11	&	0.05	&	$-$1.45	&	\nodata	&	$-$1.45	&	0.11	&	$-$0.12	&	0.06	&	0.59	&	0.10	&	0.95	&	0.14	\\
15461025$-$3746168	&	$-$0.07	&	0.06	&	$-$1.44	&	\nodata	&	$-$1.45	&	0.13	&	$-$0.22	&	0.06	&	0.48	&	0.12	&	0.80	&	0.10	\\
15461222$-$3747118	&	0.18	&	0.06	&	$-$1.48	&	0.10	&	$-$1.49	&	0.13	&	$-$0.13	&	0.07	&	0.49	&	0.11	&	0.80	&	0.18	\\
15461303$-$3746009	&	0.05	&	0.06	&	$-$1.49	&	0.10	&	$-$1.49	&	0.12	&	$-$0.03	&	0.07	&	0.59	&	0.10	&	0.79	&	0.12	\\
15461630$-$3744468	&	0.09	&	0.06	&	$-$1.53	&	0.10	&	$-$1.53	&	0.12	&	$-$0.11	&	0.05	&	0.42	&	0.10	&	0.77	&	0.10	\\
\enddata

\end{deluxetable}

\clearpage
\tablenum{5}
\tablecolumns{2}
\tablewidth{0pt}

\begin{deluxetable}{ll}
\tablecaption{Globular Cluster Literature References}
\tablehead{
\colhead{Cluster}	&
\colhead{Source} 
}

\startdata
HP-1	&	Barbuy et al. (2006)	\\
HP-1	&	Barbuy et al. (2016)	\\
M 10	&	Haynes et al. (2008)	\\
M 107	&	O'Connell et al. (2011)	\\
M 12	&	Johnson \& Pilachowski (2006)	\\
M 13	&	Sneden et al. (2004)	\\
M 13	&	Cohen \& Mel{\'e}ndez (2005)	\\
M 15	&	Sneden et al. (1997)	\\
M 2	&	Yong et al. (2014a)	\\
M 22	&	Marino et al. (2011a)	\\
M 3	&	Sneden et al. (2004)	\\
M 3	&	Cohen \& Mel{\'e}ndez (2005)	\\
M 4	&	Ivans et al. (1999)	\\
M 5	&	Ivans et al. (2001)	\\
M 62	&	Yong et al. (2014b)	\\
M 71	&	Ram{\'{\i}}rez \& Cohen(2002)	\\
NGC 6342	&	Johnson et al. (2016b)	\\
NGC 6366	&	Johnson et al. (2016b)	\\
NGC 104	&	Carretta et al. (2004)	\\
NGC 104	&	Cordero et al. (2014)	\\
NGC 1851	&	Yong \& Grundahl (2008)	\\
NGC 1851	&	Carretta et al. (2011)	\\
NGC 1851	&	Gratton et al. (2012b)	\\
NGC 1904	&	Gratton \& Ortolani (1989)	\\
NGC 2419	&	Cohen \& Kirby (2012)	\\
NGC 2808	&	Carretta (2015)	\\
NGC 288	&	Shetrone \& Keane (2000)	\\
NGC 3201	&	Gratton \& Ortolani (1989)	\\
NGC 362	&	Shetrone \& Keane (2000)	\\
NGC 362	&	Carretta et al. (2013a)	\\
NGC 4590	&	Gratton \& Ortolani (1989)	\\
NGC 4833	&	Carretta et al. (2014)	\\
NGC 4833	&	Roederer \& Thompson (2015)	\\
NGC 5286	&	Marino et al. (2015)	\\
NGC 5824	&	Roederer et al. (2016)	\\
NGC 5897	&	Gratton (1987)	\\
NGC 6093	&	Carretta et al. (2015)	\\
NGC 6273	&	Johnson et al. (2015a)	\\
NGC 6273	&	Johnson et al. (2017)  \\
NGC 6287	&	Lee \& Carney (2002)	\\
NGC 6293	&	Lee \& Carney (2002)	\\
NGC 6352	&	Feltzing et al. (2009)	\\
NGC 6362	&	Gratton (1987)	\\
NGC 6388	&	Carretta et al. (2007)	\\
NGC 6397	&	Gratton \& Ortolani (1989)	\\
NGC 6441	&	Gratton et al. (2006)	\\
NGC 6541	&	Lee \& Carney (2002)	\\
NGC 6752	&	Yong et al. (2005)	\\
\enddata

\end{deluxetable}

\clearpage
\tablenum{6}
\tablecolumns{10}
\tablewidth{0pt}

\begin{deluxetable}{lccccccccc}
\tabletypesize{\tiny}
\tablecaption{The \emph{t} statistic, Degrees of Freedom, and \emph{p}--value 
Results from the Welch's \emph{t}--test}
\tablehead{
\colhead{Populations}	&
\colhead{[O/Fe]}	&
\colhead{[Na/Fe]}      &
\colhead{[Mg/Fe]}      &
\colhead{[Al/Fe]}      &
\colhead{[Si/Fe]}      &
\colhead{[O/Na]}	&
\colhead{[Na/Mg]}        &
\colhead{[Mg/Al]}        &
\colhead{[Al/Si]}       
}

\startdata
\emph{t} (P1$-$P2)	&	1.15	&	$-$7.98	&	1.65	&	$-$3.54	&	0.33	&	6.60	&	$-$8.51	&	3.71	&	$-$4.04	\\
d.o.f.	&	8	&	6	&	10	&	7	&	7	&	7	&	12	&	10	&	12	\\
\emph{p}	&	2.83$\times$10$^{-1}$	&	1.52$\times$10$^{-4}$	&	1.29$\times$10$^{-1}$	&	8.56$\times$10$^{-3}$	&	7.48$\times$10$^{-1}$	&	2.81$\times$10$^{-4}$	&	2.56$\times$10$^{-6}$	&	3.96$\times$10$^{-3}$	&	1.57$\times$10$^{-3}$	\\
\hline
\emph{t} (P1$-$I1)	&	6.42	&	$-$11.27	&	2.32	&	$-$4.86	&	$-$0.09	&	14.85	&	$-$11.40	&	5.20	&	$-$5.70	\\
d.o.f.	&	6	&	7	&	10	&	10	&	9	&	5	&	10	&	10	&	10	\\
\emph{p}	&	5.93$\times$10$^{-4}$	&	9.40$\times$10$^{-6}$	&	4.31$\times$10$^{-2}$	&	7.33$\times$10$^{-4}$	&	9.33$\times$10$^{-1}$	&	1.69$\times$10$^{-5}$	&	5.51$\times$10$^{-7}$	&	4.22$\times$10$^{-4}$	&	2.10$\times$10$^{-4}$	\\
\hline
\emph{t} (P1$-$I2)	&	9.35	&	$-$7.89	&	9.49	&	$-$11.11	&	$-$1.56	&	17.65	&	$-$10.51	&	14.38	&	$-$15.77	\\
d.o.f.	&	4	&	2	&	5	&	5	&	4	&	4	&	1	&	4	&	4	\\
\emph{p}	&	5.45$\times$10$^{-4}$	&	2.54$\times$10$^{-2}$	&	2.68$\times$10$^{-4}$	&	1.10$\times$10$^{-4}$	&	1.92$\times$10$^{-1}$	&	2.95$\times$10$^{-5}$	&	3.53$\times$10$^{-2}$	&	7.44$\times$10$^{-5}$	&	8.53$\times$10$^{-5}$	\\
\hline
\emph{t} (P2$-$I1)	&	6.53	&	$-$5.17	&	0.83	&	$-$2.26	&	$-$0.46	&	11.03	&	$-$3.62	&	2.30	&	$-$1.95	\\
d.o.f.	&	12	&	13	&	13	&	11	&	11	&	12	&	13	&	11	&	13	\\
\emph{p}	&	2.44$\times$10$^{-5}$	&	1.71$\times$10$^{-4}$	&	4.22$\times$10$^{-1}$	&	4.61$\times$10$^{-2}$	&	6.54$\times$10$^{-1}$	&	1.76$\times$10$^{-7}$	&	3.23$\times$10$^{-3}$	&	4.22$\times$10$^{-2}$	&	7.29$\times$10$^{-2}$	\\
\hline
\emph{t} (P2$-$I2)	&	9.84	&	$-$3.82	&	8.09	&	$-$10.18	&	$-$2.41	&	14.62	&	$-$6.05	&	10.77	&	$-$8.21	\\
d.o.f.	&	3	&	1	&	6	&	7	&	2	&	3	&	1	&	5	&	9	\\
\emph{p}	&	1.65$\times$10$^{-3}$	&	1.32$\times$10$^{-1}$	&	1.47$\times$10$^{-4}$	&	2.69$\times$10$^{-5}$	&	1.16$\times$10$^{-1}$	&	9.79$\times$10$^{-4}$	&	6.02$\times$10$^{-2}$	&	1.00$\times$10$^{-4}$	&	1.68$\times$10$^{-5}$	\\
\hline
\emph{t} (I1$-$I2)	&	4.86	&	$-$1.53	&	6.24	&	$-$4.68	&	$-$1.53	&	7.09	&	$-$3.77	&	5.65	&	$-$4.36	\\
d.o.f.	&	2	&	1	&	6	&	7	&	4	&	2	&	2	&	7	&	6	\\
\emph{p}	&	3.62$\times$10$^{-2}$	&	3.32$\times$10$^{-1}$	&	6.08$\times$10$^{-4}$	&	2.26$\times$10$^{-3}$	&	1.95$\times$10$^{-1}$	&	3.14$\times$10$^{-2}$	&	9.43$\times$10$^{-2}$	&	9.01$\times$10$^{-4}$	&	4.67$\times$10$^{-3}$	\\
\enddata

\tablecomments{The ``E" population is excluded because it consists of only one 
star.}

\end{deluxetable}



\begin{thebibliography}{}

\bibitem[Allen et al.(2008)]{2008ApJ...674..237A} Allen, C., Moreno, E., \& Pichardo, B.\ 2008, \apj, 674, 237-246

\bibitem[Alves et al.(2001)]{2001AJ....121..318A} Alves, D.~R., Bond, H.~E., \& Onken, C.\ 2001, \aj, 121, 318

\bibitem[Aoki et al.(2007)]{2007PASJ...59L..15A} Aoki, W., Honda, S., Sadakane, K., \& Arimoto, N.\ 2007, \pasj, 59, L15

\bibitem[Arnould et al.(1999)]{1999A&A...347..572A} Arnould, M., Goriely, S., \& Jorissen, A.\ 1999, \aap, 347, 572

\bibitem[Barbuy et al.(2006)]{2006A&A...449..349B} Barbuy, B., Zoccali, M., Ortolani, S., et al.\ 2006, \aap, 449, 349

\bibitem[Barbuy et al.(2016)]{2016A&A...591A..53B} Barbuy, B., Cantelli, E., Vemado, A., et al.\ 2016, \aap, 591, A53 

\bibitem[Barklem et al.(2005)]{2005A&A...439..129B} Barklem, P.~S., Christlieb, N., Beers, T.~C., et al.\ 2005, \aap, 439, 129

\bibitem[Bastian et al.(2015)]{2015MNRAS.449.3333B} Bastian, N., Cabrera-Ziri, I., \& Salaris, M.\ 2015, \mnras, 449, 3333

\bibitem[Bastian \& Lardo(2015)]{2015MNRAS.453..357B} Bastian, N., \& Lardo, C.\ 2015, \mnras, 453, 357

\bibitem[Battistini \& Bensby(2016)]{2016A&A...586A..49B} Battistini, C., \& Bensby, T.\ 2016, \aap, 586, A49

\bibitem[Bekki \& Freeman(2003)]{2003MNRAS.346L..11B} Bekki, K., \& Freeman, K.~C.\ 2003, \mnras, 346, L11

\bibitem[Bica \& Pastoriza(1983)]{1983Ap&SS..91...99B} Bica, E.~L.~D., \& Pastoriza, M.~G.\ 1983, \apss, 91, 99 

\bibitem[Bragaglia et al.(2010a)]{2010ApJ...720L..41B} Bragaglia, A., Carretta, E., Gratton, R.~G., et al.\ 2010a, \apjl, 720, L41

\bibitem[Bragaglia et al.(2010b)]{2010A&A...519A..60B} Bragaglia, A., Carretta, E., Gratton, R., et al.\ 2010b, \aap, 519, A60

\bibitem[Briley et al.(1994)]{1994AJ....108.2183B} Briley, M.~M., Hesser, J.~E., Bell, R.~A., Bolte, M., \& Smith, G.~H.\ 1994, \aj, 108, 2183

\bibitem[Briley et al.(1996)]{1996Natur.383..604B} Briley, M.~M., Smith, V.~V., Suntzeff, N.~B., et al.\ 1996, \nat, 383, 604

\bibitem[Carretta et al.(2004)]{2004A&A...416..925C} Carretta, E., Gratton, R.~G., Bragaglia, A., Bonifacio, P., \& Pasquini, L.\ 2004, \aap, 416, 925

\bibitem[Carretta et al.(2007)]{2007A&A...464..967C} Carretta, E., Bragaglia, A., Gratton, R.~G., et al.\ 2007, \aap, 464, 967

\bibitem[Carretta et al.(2009a)]{2009A&A...505..117C} Carretta, E., Bragaglia, A., Gratton, R.~G., et al.\ 2009a, \aap, 505, 117

\bibitem[Carretta et al.(2009b)]{2009A&A...505..139C} Carretta, E., Bragaglia, A., Gratton, R., \& Lucatello, S.\ 2009b, \aap, 505, 139

\bibitem[Carretta et al.(2010a)]{2010A&A...516A..55C} Carretta, E., Bragaglia, A., Gratton, R.~G., et al.\ 2010a, \aap, 516, A55

\bibitem[Carretta et al.(2010b)]{2010A&A...520A..95C} Carretta, E., Bragaglia, A., Gratton, R.~G., et al.\ 2010b, \aap, 520, A95

\bibitem[Carretta et al.(2010c)]{2010A&A...519A..71C} Carretta, E., Bragaglia, A., D'Orazi, V., Lucatello, S., \& Gratton, R.~G.\ 2010c, \aap, 519, A71

\bibitem[Carretta et al.(2011)]{2011A&A...533A..69C} Carretta, E., Lucatello, S., Gratton, R.~G., Bragaglia, A., \& D'Orazi, V.\ 2011, \aap, 533, A69

\bibitem[Carretta et al.(2012a)]{2012ApJ...750L..14C} Carretta, E., Bragaglia, A., Gratton, R.~G., Lucatello, S., \& D'Orazi, V.\ 2012a, \apjl, 750, L14

\bibitem[Carretta et al.(2012b)]{2012A&A...543A.117C} Carretta, E., D'Orazi, V., Gratton, R.~G., \& Lucatello, S.\ 2012b, \aap, 543, A117

\bibitem[Carretta et al.(2013a)]{2013A&A...557A.138C} Carretta, E., Bragaglia, A., Gratton, R.~G., et al.\ 2013a, \aap, 557, A138

\bibitem[Carretta et al.(2013b)]{2013ApJ...769...40C} Carretta, E., Gratton, R.~G., Bragaglia, A., et al.\ 2013b, \apj, 769, 40

\bibitem[Carretta(2014)]{2014ApJ...795L..28C} Carretta, E.\ 2014, \apjl, 795, L28

\bibitem[Carretta et al.(2014)]{2014A&A...564A..60C} Carretta, E., Bragaglia, A., Gratton, R.~G., et al.\ 2014, \aap, 564, A60

\bibitem[Carretta(2015)]{2015ApJ...810..148C} Carretta, E.\ 2015, \apj, 810, 148

\bibitem[Carretta et al.(2015)]{2015A&A...578A.116C} Carretta, E., Bragaglia, A., Gratton, R.~G., et al.\ 2015, \aap, 578, A116

\bibitem[Casetti-Dinescu et al.(2007)]{2007AJ....134..195C} Casetti-Dinescu, D.~I., Girard, T.~M., Herrera, D., et al.\ 2007, \aj, 134, 195

\bibitem[Castelli \& Kurucz(2004)]{2004astro.ph..5087C} Castelli, F., \& Kurucz, R.~L.\ 2004, arXiv:astro-ph/0405087

\bibitem[Cavallo \& Nagar(2000)]{2000AJ....120.1364C} Cavallo, R.~M., \& Nagar, N.~M.\ 2000, \aj, 120, 1364

\bibitem[Cavallo et al.(2004)]{2004AJ....127.3411C} Cavallo, R.~M., Suntzeff, N.~B., \& Pilachowski, C.~A.\ 2004, \aj, 127, 3411

\bibitem[Cohen(1978)]{1978ApJ...223..487C} Cohen, J.~G.\ 1978, \apj, 223, 487

\bibitem[Cohen \& Mel{\'e}ndez(2005)]{2005AJ....129..303C} Cohen, J.~G., \& Mel{\'e}ndez, J.\ 2005, \aj, 129, 303

\bibitem[Cohen \& Huang(2009)]{2009ApJ...701.1053C} Cohen, J.~G., \& Huang, W.\ 2009, \apj, 701, 1053

\bibitem[Cohen \& Huang(2010)]{2010ApJ...719..931C} Cohen, J.~G., \& Huang, W.\ 2010, \apj, 719, 931

\bibitem[Cohen \& Kirby(2012)]{2012ApJ...760...86C} Cohen, J.~G., \& Kirby, E.~N.\ 2012, \apj, 760, 86

\bibitem[Cordero et al.(2014)]{2014ApJ...780...94C} Cordero, M.~J., Pilachowski, C.~A., Johnson, C.~I., et al.\ 2014, \apj, 780, 94

\bibitem[Cordero et al.(2015)]{2015ApJ...800....3C} Cordero, M.~J., Pilachowski, C.~A., Johnson, C.~I., \& Vesperini, E.\ 2015, \apj, 800, 3

\bibitem[Cordero et al.(2017)]{2017MNRAS.465.3515C} Cordero, M.~J., H{\'e}nault-Brunet, V., Pilachowski, C.~A., et al.\ 2017, \mnras, 465, 3515

\bibitem[D'Antona \& Ventura(2007)]{2007MNRAS.379.1431D} D'Antona, F., \& Ventura, P.\ 2007, \mnras, 379, 1431

\bibitem[D'Antona et al.(2016)]{2016MNRAS.458.2122D} D'Antona, F., Vesperini, E., D'Ercole, A., et al.\ 2016, \mnras, 458, 2122

\bibitem[D'Ercole et al.(2008)]{2008MNRAS.391..825D} D'Ercole, A., Vesperini, E., D'Antona, F., McMillan, S.~L.~W., \& Recchi, S.\ 2008, \mnras, 391, 825

\bibitem[D'Orazi et al.(2010)]{2010ApJ...713L...1D} D'Orazi, V., Lucatello, S., Gratton, R., et al.\ 2010, \apjl, 713, L1

\bibitem[Da Costa et al.(2013)]{2013ApJ...769....8D} Da Costa, G.~S., Norris, J.~E., \& Yong, D.\ 2013, \apj, 769, 8

\bibitem[da Costa(2016)]{2016IAUS..317..110D} da Costa, G.~S.\ 2016, The General Assembly of Galaxy Halos: Structure, Origin and Evolution, 317, 110

\bibitem[Dalessandro et al.(2014)]{2014ApJ...791L...4D} Dalessandro, E., Massari, D., Bellazzini, M., et al.\ 2014, \apjl, 791, L4

\bibitem[Dalessandro et al.(2016)]{2016ApJ...829...77D} Dalessandro, E., Lapenna, E., Mucciarelli, A., et al.\ 2016, \apj, 829, 77

\bibitem[Decressin et al.(2007)]{2007A&A...464.1029D} Decressin, T., Meynet, G., Charbonnel, C., Prantzos, N., \& Ekstr{\"o}m, S.\ 2007, \aap, 464, 1029

\bibitem[de Mink et al.(2009)]{2009A&A...507L...1D} de Mink, S.~E., Pols, O.~R., Langer, N., \& Izzard, R.~G.\ 2009, \aap, 507, L1

\bibitem[Denisenkov \& Denisenkova(1990)]{1990SvAL...16..275D} Denisenkov, P.~A., \& Denisenkova, S.~N.\ 1990, Soviet Astronomy Letters, 16, 275

\bibitem[Denissenkov \& Hartwick(2014)]{2014MNRAS.437L..21D} Denissenkov, P.~A., \& Hartwick, F.~D.~A.\ 2014, \mnras, 437, L21 

\bibitem[Dobrovolskas et al.(2014)]{2014A&A...565A.121D} Dobrovolskas, V., Ku{\v c}inskas, A., Bonifacio, P., et al.\ 2014, \aap, 565, A121

\bibitem[Doherty et al.(2014)]{2014MNRAS.441..582D} Doherty, C.~L., Gil-Pons, P., Lau, H.~H.~B., et al.\ 2014, \mnras, 441, 582

\bibitem[Dotter et al.(2010)]{2010ApJ...708..698D} Dotter, A., Sarajedini, A., Anderson, J., et al.\ 2010, \apj, 708, 698

\bibitem[Drake et al.(1992)]{1992ApJ...395L..95D} Drake, J.~J., Smith, V.~V., \& Suntzeff, N.~B.\ 1992, \apjl, 395, L95

\bibitem[Dupree et al.(2011)]{2011ApJ...728..155D} Dupree, A.~K., Strader, J., \& Smith, G.~H.\ 2011, \apj, 728, 155

\bibitem[Dupree et al.(2016)]{2016ApJ...821L...7D} Dupree, A.~K., Avrett, E.~H., \& Kurucz, R.~L.\ 2016, \apjl, 821, L7

\bibitem[Feltzing et al.(2009)]{2009A&A...493..913F} Feltzing, S., Primas, F., \& Johnson, R.~A.\ 2009, \aap, 493, 913

\bibitem[Geisler et al.(1997)]{1997PASP..109..799G} Geisler, D., Claria, J.~J., \& Minniti, D.\ 1997, \pasp, 109, 799 

\bibitem[Georgiev et al.(2009)]{2009MNRAS.396.1075G} Georgiev, I.~Y., Hilker, M., Puzia, T.~H., Goudfrooij, P., \& Baumgardt, H.\ 2009, \mnras, 396, 1075

\bibitem[Gratton(1987)]{1987A&A...179..181G} Gratton, R.~G.\ 1987, \aap, 179, 181

\bibitem[Gratton \& Ortolani(1989)]{1989A&A...211...41G} Gratton, R.~G., \& Ortolani, S.\ 1989, \aap, 211, 41

\bibitem[Gratton et al.(2001)]{2001A&A...369...87G} Gratton, R.~G., Bonifacio, P., Bragaglia, A., et al.\ 2001, \aap, 369, 87

\bibitem[Gratton et al.(2004)]{2004ARA&A..42..385G} Gratton, R., Sneden, C., \& Carretta, E.\ 2004, \araa, 42, 385

\bibitem[Gratton et al.(2006)]{2006A&A...455..271G} Gratton, R.~G., Lucatello, S., Bragaglia, A., et al.\ 2006, \aap, 455, 271

\bibitem[Gratton et al.(2011)]{2011A&A...534A..72G} Gratton, R.~G., Johnson, C.~I., Lucatello, S., D'Orazi, V., \& Pilachowski, C.\ 2011, \aap, 534, A72

\bibitem[Gratton et al.(2012a)]{2012A&ARv..20...50G} Gratton, R.~G., Carretta, E., \& Bragaglia, A.\ 2012a, \aapr, 20, 50

\bibitem[Gratton et al.(2012b)]{2012A&A...539A..19G} Gratton, R.~G., Lucatello, S., Carretta, E., et al.\ 2012b, \aap, 539, A19

\bibitem[Hansen et al.(2017)]{2017ApJ...838...44H} Hansen, T.~T., Simon, J.~D., Marshall, J.~L., et al.\ 2017, \apj, 838, 44

\bibitem[Harris(1996)]{1996AJ....112.1487H} Harris, W.~E.\ 1996, \aj, 112, 1487

\bibitem[Hatzes(1987)]{1987PASP...99..369H} Hatzes, A.~P.\ 1987, \pasp, 99, 369

\bibitem[Haynes et al.(2008)]{2008PASP..120.1097H} Haynes, S., Burks, G., Johnson, C.~I., \& Pilachowski, C.~A.\ 2008, \pasp, 120, 1097

\bibitem[Hesser et al.(1986)]{1986PASP...98..403H} Hesser, J.~E., Shawl, S.~J., \& Meyer, J.~E.\ 1986, \pasp, 98, 403

\bibitem[Hollyhead et al.(2017)]{2017MNRAS.465L..39H} Hollyhead, K., Kacharov, N., Lardo, C., et al.\ 2017, \mnras, 465, L39

\bibitem[Iannicola et al.(2009)]{2009ApJ...696L.120I} Iannicola, G., Monelli, M., Bono, G., et al.\ 2009, \apjl, 696, L120

\bibitem[Ivans et al.(1999)]{1999AJ....118.1273I} Ivans, I.~I., Sneden, C., Kraft, R.~P., et al.\ 1999, \aj, 118, 1273

\bibitem[Ivans et al.(2001)]{2001AJ....122.1438I} Ivans, I.~I., Kraft, R.~P., Sneden, C., et al.\ 2001, \aj, 122, 1438

\bibitem[Izzard et al.(2007)]{2007A&A...466..641I} Izzard, R.~G., Lugaro, M., Karakas, A.~I., Iliadis, C., \& van Raai, M.\ 2007, \aap, 466, 641

\bibitem[Jasniewicz et al.(2004)]{2004A&A...423..353J} Jasniewicz, G., de Laverny, P., Parthasarathy, M., L{\`e}bre, A., \& Th{\'e}venin, F.\ 2004, \aap, 423, 353

\bibitem[Ji et al.(2016)]{2016ApJ...830...93J} Ji, A.~P., Frebel, A., Simon, J.~D., \& Chiti, A.\ 2016, \apj, 830, 93

\bibitem[Johnson et al.(2005)]{2005PASP..117.1308J} Johnson, C.~I., Kraft, R.~P., Pilachowski, C.~A., et al.\ 2005, \pasp, 117, 1308 

\bibitem[Johnson \& Pilachowski(2006)]{2006AJ....132.2346J} Johnson, C.~I., \& Pilachowski, C.~A.\ 2006, \aj, 132, 2346 

\bibitem[Johnson \& Pilachowski(2010)]{2010ApJ...722.1373J} Johnson, C.~I., \& Pilachowski, C.~A.\ 2010, \apj, 722, 1373

\bibitem[Johnson \& Pilachowski(2012)]{2012ApJ...754L..38J} Johnson, C.~I., \& Pilachowski, C.~A.\ 2012, \apjl, 754, L38

\bibitem[Johnson et al.(2012)]{2012ApJ...749..175J} Johnson, C.~I., Rich, R.~M., Kobayashi, C., \& Fulbright, J.~P.\ 2012, \apj, 749, 175

\bibitem[Johnson et al.(2013)]{2013ApJ...775L..27J} Johnson, C.~I., McWilliam, A., \& Rich, R.~M.\ 2013, \apjl, 775, L27

\bibitem[Johnson et al.(2014)]{2014AJ....148...67J} Johnson, C.~I., Rich, R.~M., Kobayashi, C., Kunder, A., \& Koch, A.\ 2014, \aj, 148, 67

\bibitem[Johnson et al.(2015a)]{2015AJ....150...63J} Johnson, C.~I., Rich, R.~M., Pilachowski, C.~A., et al.\ 2015a, \aj, 150, 63

\bibitem[Johnson et al.(2015b)]{2015AJ....149...71J} Johnson, C.~I., McDonald, I., Pilachowski, C.~A., et al.\ 2015b, \aj, 149, 71

\bibitem[Johnson et al.(2016)]{2016AJ....152...21J} Johnson, C.~I., Caldwell, N., Rich, R.~M., Pilachowski, C.~A., \& Hsyu, T.\ 2016, \aj, 152, 21

\bibitem[Johnson et al.(2017)]{2017ApJ...836..168J} Johnson, C.~I., Caldwell, N., Rich, R.~M., et al.\ 2017, \apj, 836, 168

\bibitem[Kraft et al.(1993)]{1993AJ....106.1490K} Kraft, R.~P., Sneden, C., Langer, G.~E., \& Shetrone, M.~D.\ 1993, \aj, 106, 1490

\bibitem[Kraft(1994)]{1994PASP..106..553K} Kraft, R.~P.\ 1994, \pasp, 106, 553

\bibitem[Kraft et al.(1997)]{1997AJ....113..279K} Kraft, R.~P., Sneden, C., Smith, G.~H., et al.\ 1997, \aj, 113, 279

\bibitem[Kraft \& Ivans(2003)]{2003PASP..115..143K} Kraft, R.~P., \& Ivans, I.~I.\ 2003, \pasp, 115, 143

\bibitem[Kravtsov et al.(1997)]{1997AstL...23..391K} Kravtsov, V.~V., Pavlov, M.~V., Samus', N.~N., et al.\ 1997, Astronomy Letters, 23, 391

\bibitem[Kravtsov et al.(2011)]{2011A&A...527L...9K} Kravtsov, V., Alca{\'{\i}}no, G., Marconi, G., \& Alvarado, F.\ 2011, \aap, 527, L9

\bibitem[Kurtz \& Mink(1998)]{1998PASP..110..934K} Kurtz, M.~J., \& Mink, D.~J.\ 1998, \pasp, 110, 934

\bibitem[Langer et al.(1993)]{1993PASP..105..301L} Langer, G.~E., Hoffman, R., \& Sneden, C.\ 1993, \pasp, 105, 301

\bibitem[Langer et al.(1997)]{1997PASP..109..244L} Langer, G.~E., Hoffman, R.~E., \& Zaidins, C.~S.\ 1997, \pasp, 109, 244

\bibitem[Lardo et al.(2011)]{2011A&A...525A.114L} Lardo, C., Bellazzini, M., Pancino, E., et al.\ 2011, \aap, 525, A114

\bibitem[Lardo et al.(2016)]{2016MNRAS.457...51L} Lardo, C., Mucciarelli, A., \& Bastian, N.\ 2016, \mnras, 457, 51

\bibitem[Larsen et al.(2014)]{2014ApJ...797...15L} Larsen, S.~S., Brodie, J.~P., Grundahl, F., \& Strader, J.\ 2014, \apj, 797, 15

\bibitem[Larsen et al.(2015)]{2015ApJ...804...71L} Larsen, S.~S., Baumgardt, H., Bastian, N., et al.\ 2015, \apj, 804, 71

\bibitem[Lawler et al.(2001a)]{2001ApJ...556..452L} Lawler, J.~E., Bonvallet, G., \& Sneden, C.\ 2001a, \apj, 556, 452

\bibitem[Lawler et al.(2001b)]{2001ApJ...563.1075L} Lawler, J.~E., Wickliffe, M.~E., den Hartog, E.~A., \& Sneden, C.\ 2001b, \apj, 563, 1075

\bibitem[Lee \& Carney(2002)]{2002AJ....124.1511L} Lee, J.-W., \& Carney, B.~W.\ 2002, \aj, 124, 1511

\bibitem[Lee et al.(2007)]{2007ApJ...661L..49L} Lee, Y.-W., Gim, H.~B., \& Casetti-Dinescu, D.~I.\ 2007, \apjl, 661, L49

\bibitem[Lee(2016)]{2016ApJS..226...16L} Lee, J.-W.\ 2016, \apjs, 226, 16

\bibitem[Lemasle et al.(2014)]{2014A&A...572A..88L} Lemasle, B., de Boer, T.~J.~L., Hill, V., et al.\ 2014, \aap, 572, A88

\bibitem[Letarte et al.(2010)]{2010A&A...523A..17L} Letarte, B., Hill, V., Tolstoy, E., et al.\ 2010, \aap, 523, A17

\bibitem[Lim et al.(2016)]{2016ApJ...832...99L} Lim, D., Lee, Y.-W., Pasquato, M., Han, S.-I., \& Roh, D.-G.\ 2016, \apj, 832, 99

\bibitem[Marino et al.(2009)]{2009A&A...505.1099M} Marino, A.~F., Milone, A.~P., Piotto, G., et al.\ 2009, \aap, 505, 1099

\bibitem[Marino et al.(2011a)]{2011A&A...532A...8M} Marino, A.~F., Sneden, C., Kraft, R.~P., et al.\ 2011a, \aap, 532, A8

\bibitem[Marino et al.(2011b)]{2011ApJ...731...64M} Marino, A.~F., Milone, A.~P., Piotto, G., et al.\ 2011b, \apj, 731, 64  

\bibitem[Marino et al.(2015)]{2015MNRAS.450..815M} Marino, A.~F., Milone, A.~P., Karakas, A.~I., et al.\ 2015, \mnras, 450, 815 

\bibitem[Massari et al.(2016)]{2016MNRAS.458.4162M} Massari, D., Lapenna, E., Bragaglia, A., et al.\ 2016, \mnras, 458, 4162

\bibitem[Mateo et al.(2012)]{2012SPIE.8446E..4YM} Mateo, M., Bailey, J.~I., Crane, J., et al.\ 2012, \procspie, 8446, 84464Y

\bibitem[McWilliam et al.(2010)]{2010IAUS..265..279M} McWilliam, A., Fulbright, J., \& Rich, R.~M.\ 2010, Chemical Abundances in the Universe: Connecting First Stars to Planets, 265, 279

\bibitem[M{\'e}sz{\'a}ros et al.(2015)]{2015AJ....149..153M} M{\'e}sz{\'a}ros, S., Martell, S.~L., Shetrone, M., et al.\ 2015, \aj, 149, 153

\bibitem[Miholics et al.(2015)]{2015MNRAS.454.2166M} Miholics, M., Webb, J.~J., \& Sills, A.\ 2015, \mnras, 454, 2166

\bibitem[Milone et al.(2012a)]{2012ApJ...744...58M} Milone, A.~P., Piotto, G., Bedin, L.~R., et al.\ 2012a, \apj, 744, 58

\bibitem[Milone et al.(2012b)]{2012ApJ...745...27M} Milone, A.~P., Marino, A.~F., Piotto, G., et al.\ 2012b, \apj, 745, 27

\bibitem[Milone et al.(2013)]{2013ApJ...767..120M} Milone, A.~P., Marino, A.~F., Piotto, G., et al.\ 2013, \apj, 767, 120

\bibitem[Milone(2015)]{2015MNRAS.446.1672M} Milone, A.~P.\ 2015, \mnras, 446, 1672

\bibitem[Milone et al.(2015a)]{2015MNRAS.447..927M} Milone, A.~P., Marino, A.~F., Piotto, G., et al.\ 2015a, \mnras, 447, 927

\bibitem[Milone et al.(2015b)]{2015ApJ...808...51M} Milone, A.~P., Marino, A.~F., Piotto, G., et al.\ 2015b, \apj, 808, 51

\bibitem[Milone et al.(2017)]{2017MNRAS.464.3636M} Milone, A.~P., Piotto, G., Renzini, A., et al.\ 2017, \mnras, 464, 3636

\bibitem[Momany et al.(2004)]{2004A&A...420..605M} Momany, Y., Bedin, L.~R., Cassisi, S., et al.\ 2004, \aap, 420, 605

\bibitem[Monelli et al.(2013)]{2013MNRAS.431.2126M} Monelli, M., Milone, A.~P., Stetson, P.~B., et al.\ 2013, \mnras, 431, 2126

\bibitem[Moni Bidin et al.(2009)]{2009A&A...498..737M} Moni Bidin, C., Moehler, S., Piotto, G., Momany, Y., \& Recio-Blanco, A.\ 2009, \aap, 498, 737

\bibitem[Moreno et al.(2014)]{2014ApJ...793..110M} Moreno, E., Pichardo, B., \& Vel{\'a}zquez, H.\ 2014, \apj, 793, 110

\bibitem[Mottini et al.(2008)]{2008AJ....136..614M} Mottini, M., Wallerstein, G., \& McWilliam, A.\ 2008, \aj, 136, 614

\bibitem[Mucciarelli et al.(2009)]{2009ApJ...695L.134M} Mucciarelli, A., Origlia, L., Ferraro, F.~R., \& Pancino, E.\ 2009, \apjl, 695, L134

\bibitem[Mucciarelli et al.(2012)]{2012MNRAS.426.2889M} Mucciarelli, A., Bellazzini, M., Ibata, R., et al.\ 2012, \mnras, 426, 2889

\bibitem[Mucciarelli et al.(2015a)]{2015ApJ...801...69M} Mucciarelli, A., Lapenna, E., Massari, D., Ferraro, F.~R., \& Lanzoni, B.\ 2015a, \apj, 801, 69

\bibitem[Mucciarelli et al.(2015b)]{2015ApJ...809..128M} Mucciarelli, A., Lapenna, E., Massari, D., et al.\ 2015b, \apj, 809, 128

\bibitem[Mucciarelli et al.(2015c)]{2015ApJ...801...68M} Mucciarelli, A., Bellazzini, M., Merle, T., et al.\ 2015c, \apj, 801, 68

\bibitem[Nardiello et al.(2015)]{2015A&A...573A..70N} Nardiello, D., Milone, A.~P., Piotto, G., et al.\ 2015, \aap, 573, A70

\bibitem[Nataf et al.(2011)]{2011ApJ...736...94N} Nataf, D.~M., Gould, A., Pinsonneault, M.~H., \& Stetson, P.~B.\ 2011, \apj, 736, 94

\bibitem[Niederhofer et al.(2017)]{2017MNRAS.465.4159N} Niederhofer, F., Bastian, N., Kozhurina-Platais, V., et al.\ 2017, \mnras, 465, 4159

\bibitem[Norris et al.(1981)]{1981ApJ...244..205N} Norris, J., Cottrell, P.~L., Freeman, K.~C., \& Da Costa, G.~S.\ 1981, \apj, 244, 205

\bibitem[Norris \& Pilachowski(1985)]{1985ApJ...299..295N} Norris, J., \& Pilachowski, C.~A.\ 1985, \apj, 299, 295 

\bibitem[Norris \& Da Costa(1995)]{1995ApJ...447..680N} Norris, J.~E., \& Da Costa, G.~S.\ 1995, \apj, 447, 680

\bibitem[Norris(2004)]{2004ApJ...612L..25N} Norris, J.~E.\ 2004, \apjl, 612, L25

\bibitem[O'Connell et al.(2011)]{2011PASP..123.1139O} O'Connell, J.~E., Johnson, C.~I., Pilachowski, C.~A., \& Burks, G.\ 2011, \pasp, 123, 1139

\bibitem[Ortolani et al.(2000)]{2000A&A...362..953O} Ortolani, S., Momany, Y., Barbuy, B., Bica, E., \& Catelan, M.\ 2000, \aap, 362, 953

\bibitem[Pasquini et al.(2011)]{2011A&A...531A..35P} Pasquini, L., Mauas, P., K{\"a}ufl, H.~U., \& Cacciari, C.\ 2011, \aap, 531, A35

\bibitem[Peterson(1980)]{1980ApJ...237L..87P} Peterson, R.~C.\ 1980, \apjl, 237, L87

\bibitem[Pilachowski et al.(1996)]{1996AJ....112..545P} Pilachowski, C.~A., Sneden, C., Kraft, R.~P., \& Langer, G.~E.\ 1996, \aj, 112, 545

\bibitem[Piotto et al.(2005)]{2005ApJ...621..777P} Piotto, G., Villanova, S., Bedin, L.~R., et al.\ 2005, \apj, 621, 777

\bibitem[Piotto et al.(2007)]{2007ApJ...661L..53P} Piotto, G., Bedin, L.~R., Anderson, J., et al.\ 2007, \apjl, 661, L53

\bibitem[Piotto et al.(2012)]{2012ApJ...760...39P} Piotto, G., Milone, A.~P., Anderson, J., et al.\ 2012, \apj, 760, 39

\bibitem[Piotto et al.(2015)]{2015AJ....149...91P} Piotto, G., Milone, A.~P., Bedin, L.~R., et al.\ 2015, \aj, 149, 91

\bibitem[Prantzos et al.(2007)]{2007A&A...470..179P} Prantzos, N., Charbonnel, C., \& Iliadis, C.\ 2007, \aap, 470, 179

\bibitem[Rakos \& Schombert(2005)]{2005PASP..117..245R} Rakos, K., \& Schombert, J.\ 2005, \pasp, 117, 245

\bibitem[Ram{\'{\i}}rez \& Cohen(2002)]{2002AJ....123.3277R} Ram{\'{\i}}rez, S.~V., \& Cohen, J.~G.\ 2002, \aj, 123, 3277

\bibitem[Recio-Blanco et al.(2005)]{2005A&A...432..851R} Recio-Blanco, A., Piotto, G., de Angeli, F., et al.\ 2005, \aap, 432, 851

\bibitem[Renzini(2008)]{2008MNRAS.391..354R} Renzini, A.\ 2008, \mnras, 391, 354

\bibitem[Renzini et al.(2015)]{2015MNRAS.454.4197R} Renzini, A., D'Antona, F., Cassisi, S., et al.\ 2015, \mnras, 454, 4197

\bibitem[Richer et al.(2013)]{2013ApJ...771L..15R} Richer, H.~B., Heyl, J., Anderson, J., et al.\ 2013, \apjl, 771, L15

\bibitem[Roederer(2011)]{2011ApJ...732L..17R} Roederer, I.~U.\ 2011, \apjl, 732, L17

\bibitem[Roederer et al.(2014)]{2014AJ....147..136R} Roederer, I.~U., Preston, G.~W., Thompson, I.~B., et al.\ 2014, \aj, 147, 136

\bibitem[Roederer \& Thompson(2015)]{2015MNRAS.449.3889R} Roederer, I.~U., \& Thompson, I.~B.\ 2015, \mnras, 449, 3889

\bibitem[Roederer et al.(2016)]{2016MNRAS.455.2417R} Roederer, I.~U., Mateo, M., Bailey, J.~I., et al.\ 2016, \mnras, 455, 2417

\bibitem[Rosenberg et al.(2000)]{2000A&AS..144....5R} Rosenberg, A., Piotto, G., Saviane, I., \& Aparicio, A.\ 2000, \aaps, 144, 5

\bibitem[Rutledge et al.(1997)]{1997PASP..109..883R} Rutledge, G.~A., Hesser, J.~E., Stetson, P.~B., et al.\ 1997, \pasp, 109, 883 

\bibitem[Sbordone et al.(2015)]{2015A&A...579A.104S} Sbordone, L., Monaco, L., Moni Bidin, C., et al.\ 2015, \aap, 579, A104

\bibitem[Schiavon et al.(2013)]{2013ApJ...776L...7S} Schiavon, R.~P., Caldwell, N., Conroy, C., et al.\ 2013, \apjl, 776, L7

\bibitem[Shetrone \& Keane(2000)]{2000AJ....119..840S} Shetrone, M.~D., \& Keane, M.~J.\ 2000, \aj, 119, 840

\bibitem[Shetrone et al.(2001)]{2001ApJ...548..592S} Shetrone, M.~D., C{\^o}t{\'e}, P., \& Sargent, W.~L.~W.\ 2001, \apj, 548, 592

\bibitem[Shetrone et al.(2003)]{2003AJ....125..684S} Shetrone, M., Venn, K.~A., Tolstoy, E., et al.\ 2003, \aj, 125, 684

\bibitem[Simioni et al.(2016)]{2016MNRAS.463..449S} Simioni, M., Milone, A.~P., Bedin, L.~R., et al.\ 2016, \mnras, 463, 449

\bibitem[Skrutskie et al.(2006)]{2006AJ....131.1163S} Skrutskie, M.~F., Cutri, R.~M., Stiening, R., et al.\ 2006, \aj, 131, 1163

\bibitem[Smith et al.(2000)]{2000AJ....119.1239S} Smith, V.~V., Suntzeff, N.~B., Cunha, K., et al.\ 2000, \aj, 119, 1239

\bibitem[Sneden(1973)]{1973ApJ...184..839S} Sneden, C.\ 1973, \apj, 184, 839

\bibitem[Sneden et al.(1997)]{1997AJ....114.1964S} Sneden, C., Kraft, R.~P., Shetrone, M.~D., et al.\ 1997, \aj, 114, 1964

\bibitem[Sneden et al.(2000)]{2000AJ....120.1351S} Sneden, C., Pilachowski, C.~A., \& Kraft, R.~P.\ 2000, \aj, 120, 1351 

\bibitem[Sneden et al.(2004)]{2004AJ....127.2162S} Sneden, C., Kraft, R.~P., Guhathakurta, P., Peterson, R.~C., \& Fulbright, J.~P.\ 2004, \aj, 127, 2162

\bibitem[Sneden et al.(2014)]{2014ApJS..214...26S} Sneden, C., Lucatello, S., Ram, R.~S., Brooke, J.~S.~A., \& Bernath, P.\ 2014, \apjs, 214, 26

\bibitem[Sobeck et al.(2011)]{2011AJ....141..175S} Sobeck, J.~S., Kraft, R.~P., Sneden, C., et al.\ 2011, \aj, 141, 175

\bibitem[Soto et al.(2017)]{2017AJ....153...19S} Soto, M., Bellini, A., Anderson, J., et al.\ 2017, \aj, 153, 19

\bibitem[Valcarce \& Catelan(2011)]{2011A&A...533A.120V} Valcarce, A.~A.~R., \& Catelan, M.\ 2011, \aap, 533, A120

\bibitem[Van der Swaelmen et al.(2016)]{2016A&A...586A...1V} Van der Swaelmen, M., Barbuy, B., Hill, V., et al.\ 2016, \aap, 586, A1

\bibitem[Vanderbeke et al.(2015)]{2015MNRAS.451..275V} Vanderbeke, J., De Propris, R., De Rijcke, S., et al.\ 2015, \mnras, 451, 275

\bibitem[Ventura \& D'Antona(2009)]{2009A&A...499..835V} Ventura, P., \& D'Antona, F.\ 2009, \aap, 499, 835

\bibitem[Ventura et al.(2012)]{2012ApJ...761L..30V} Ventura, P., D'Antona, F., Di Criscienzo, M., et al.\ 2012, \apjl, 761, L30

\bibitem[Ventura et al.(2016)]{2016ApJ...831L..17V} Ventura, P., Garc{\'{\i}}a-Hern{\'a}ndez, D.~A., Dell'Agli, F., et al.\ 2016, \apjl, 831, L17

\bibitem[Vesperini et al.(2013)]{2013MNRAS.429.1913V} Vesperini, E., McMillan, S.~L.~W., D'Antona, F., \& D'Ercole, A.\ 2013, \mnras, 429, 1913

\bibitem[Villanova et al.(2012)]{2012ApJ...748...62V} Villanova, S., Geisler, D., Piotto, G., \& Gratton, R.~G.\ 2012, \apj, 748, 62

\bibitem[Villanova et al.(2017)]{2017MNRAS.464.2730V} Villanova, S., Moni Bidin, C., Mauro, F., Munoz, C., \& Monaco, L.\ 2017, \mnras, 464, 2730

\bibitem[Yong et al.(2005)]{2005A&A...438..875Y} Yong, D., Grundahl, F., Nissen, P.~E., Jensen, H.~R., \& Lambert, D.~L.\ 2005, \aap, 438, 875

\bibitem[Yong \& Grundahl(2008)]{2008ApJ...672L..29Y} Yong, D., \& Grundahl, F.\ 2008, \apjl, 672, L29

\bibitem[Yong et al.(2014a)]{2014MNRAS.441.3396Y} Yong, D., Roederer, I.~U., Grundahl, F., et al.\ 2014a, \mnras, 441, 3396

\bibitem[Yong et al.(2014b)]{2014MNRAS.439.2638Y} Yong, D., Alves Brito, A., Da Costa, G.~S., et al.\ 2014b, \mnras, 439, 2638 

\bibitem[Zinn(1980)]{1980ApJS...42...19Z} Zinn, R.\ 1980, \apjs, 42, 19

\bibitem[Zinn \& West(1984)]{1984ApJS...55...45Z} Zinn, R., \& West, M.~J.\ 1984, \apjs, 55, 45

\end{thebibliography}
\end{document}